\documentclass[]{aa}
\usepackage{url}
\usepackage{natbib}             
\usepackage{graphicx}           
\usepackage{amssymb}            
\usepackage[usenames,dvipsnames]{color}  
\usepackage{txfonts}
\usepackage{color}
\usepackage{tabularx,ragged2e,booktabs} 
\newcolumntype{C}{>{\centering\arraybackslash}X} 
\usepackage{lipsum}

\newcommand{\degree}{^\circ} 

\newcommand{\BE}{\begin{equation}}
\newcommand{\EE}{\end{equation}}
\newcommand{\BA}{\begin{eqnarray}}
\newcommand{\EA}{\end{eqnarray}}
\newcommand{\fig}[1]{Fig.~\ref{fig_#1}}

\newcommand{\figs}[2]{Figs.~\ref{fig_#1} and \ref{fig_#2}}

\newcommand{\sect}[1]{Sect.~\ref{sect_#1}}
\newcommand{\app}[1]{Appendix~\ref{app_#1}}

\newcommand{\eq}[1]{Eq.~(\ref{eq_#1})}

\newcommand{\eg}{e.g.}

\newcommand{\ie}{i.e.}

\newcommand{\Bz}{B_{\rm z}}
\newcommand{\Cit}{C_{\rm it}}
\newcommand{\Fz}{F_{\rm z}}
\newcommand{\Fzax}{F_{\rm z}^{\rm axial}}
\newcommand{\Fzaz}{F_{\rm z}^{\rm azimuthal}}
\newcommand{\Nt}{N_{\rm t}}
\newcommand{\nit}{n_{\rm it}}
\newcommand{\phia}{\phi_{\rm a}}
\newcommand{\phii}{\phi_{\rm i}}
\newcommand{\phic}{\phi_{\rm c}}
\newcommand{\asy}{{\rm asy}}
\newcommand{\Bt}{B_{\rm \theta}}
\newcommand{\Bp}{B_{\rm \phi}}

\newcommand{\Nto}{N_{\rm t,0}}

\begin{document}

\title{Correcting the effect of magnetic tongues on the tilt angle of bipolar active regions}

\titlerunning{Correcting magnetic tongues effects}
\authorrunning{Poisson et al.}

\author{M. Poisson\inst{1}, M.C. L\'opez Fuentes\inst{1}, C.H. Mandrini\inst{1,3}, P. D\'emoulin\inst{2} \and C. MacCormack\inst{1}}
   \offprints{M. Poisson}
\institute{
\\
$^{1}$ Instituto de Astronom\'\i a y F\'\i sica del Espacio, IAFE, CONICET-UBA, CC. 67, Suc. 28, 1428 Buenos Aires, Argentina, \email{mpoisson@iafe.uba.ar, lopezf@iafe.uba.ar} \\
$^{2}$ LESIA, Observatoire de Paris, Universit\'e PSL, CNRS, Sorbonne Universit\'e, Univ. Paris Diderot, Sorbonne Paris Cit\'e, 5 place Jules Janssen, 92195 Meudon, France, \email{Pascal.Demoulin@obspm.fr}\\
$^{3}$ Universidad de Buenos Aires, Facultad de Ciencias Exactas y Naturales, 1428 Buenos Aires, Argentina, \email{mandrini@iafe.uba.ar}\\
}

   \abstract  
   {The magnetic polarities of bipolar 
   active regions (ARs) exhibit  elongations in line-of-sight 
   magnetograms during their emergence. These elongations are referred to as magnetic tongues and 
    attributed to the presence of twist in the emerging 
   magnetic flux-ropes (FRs) that form ARs.
} 
   {The presence of magnetic tongues affects the measurement of any AR characteristic that depends on its magnetic flux distribution. The AR tilt-angle is one of them.  
   We aim to develop a method to isolate and remove the flux associated with the tongues to determine  the AR tilt-angle with as much precision as possible.} 
   {As a first approach, we used a simple emergence model of a FR.
   This allowed us to develop and test our aim based on a method to remove the effects of magnetic tongues.  
   Then, using the experience gained from the analysis of the model, we applied our method to photospheric observations of bipolar ARs that show clear magnetic tongues.} 
   {Using the developed procedure on the FR model, we can reduce the deviation in the tilt estimation by more than $60\%$. 
   Next we illustrate the performance of the method with four examples of bipolar ARs selected for their large magnetic tongues.  
   The new method efficiently removes  the spurious rotation of the bipole. 
   This correction is mostly independent of the method input parameters and significant since it is larger than all the estimated tilt errors. }
   {We have developed a method to isolate the magnetic flux associated with the FR core during the emergence 
of bipolar ARs. 
   This allows us to compute the AR tilt-angle and its evolution 
as precisely as possible. We suggest that the high dispersion observed in the determination of AR tilt-angles in studies that massively compute them from line-of sight magnetograms can be partly due to the existence of magnetic tongues whose presence is not sufficiently acknowledged.}


   \maketitle

\section{Introduction} 
\label{sect_Introduction}

A dynamo mechanism located at the bottom of the convection zone (CZ) is frequently invoked to explain the formation of active regions \citep[ARs, see e.g. the reviews of][and references therein]{Charbonneau14,Brun17}. 
In this context, the magnetic flux is amplified and deformed by differential rotation and convective motions until it becomes buoyant and emerges in the form of twisted flux-tubes or flux ropes (FRs). 
Several magnetohydrodynamic (MHD) simulations of flux emergence consider the rise of coherent FRs from deep in the CZ that will later form ARs, once the rope succeeds to traverse the photosphere \citep[see the reviews by][and references therein]{Fan09,Cheung14,Toriumi14}. 
However, there are also numerical simulations that explain the formation of ARs due to the in situ amplification and structuring of the magnetic field by convection \citep[see the review in][]{Brandenburg18} 

The structure of the emerging FRs determines the observed characteristics of ARs in line-of-sight (LOS) magnetograms. One of these characteristics is the presence of magnetic tongues \citep{Lopez-Fuentes00}, 
which have been also called tails \citep{Archontis10}.
If we assume that ARs are formed by the emergence of $\Omega$-shaped FRs, magnetic tongues are produced by the projection of the azimuthal component along the LOS 
and they appear in bipolar ARs as elongations of the main polarities with a yin-yang pattern. 
\citet{Lopez-Fuentes00} reported their existence for the first time in a rotating bipolar AR and they were later observed in many other examples \citep[see e.g.][]{Luoni11,Mandrini14,Valori15,Yardley16,Vemareddy17,Dacie18,Lopez-Fuentes18}. 
They also have been found in numerical simulations of emerging FRs \citep{Archontis10,Cheung10,MacTaggart11,Jouve13,Rempel14,Takasao15}. 

In \citet{Poisson15b} we presented a systematic method to quantify the influence of magnetic tongues using the evolution of the photospheric inversion line (PIL) during the emergence of bipolar ARs. 
We measure the acute angle between the estimated PIL and the line orthogonal to the AR bipole axis. 
From this angle, which we called the tongue angle [$\tau$], we estimated the average twist present in the sub-photospheric emerging FR under the assumption that it can be modelled using a uniformly twisted half torus. 
We found that, in general, the twist is below one turn.

In a subsequent article \citep{Poisson15a}, for a set of simple
bipolar ARs, we compared  the twist computed as in \citet{Poisson15b} with the twist inferred using a linear force-free field model of the AR coronal field. The signs of the twist estimated from both methods are consistent. 
Furthermore, we found a linear relation between the twist derived from the analysis of tongues and that obtained from the coronal field model.

In \citet{Poisson16},  we studied how the tongues affect the evolution of the magnetic flux distribution of ARs. 
This study was done for bipolar ARs observed over more than one solar cycle. 
We also developed a more sophisticated FR emergence model that considered loop cross-sections
with non-uniform twists (both in the radial and azimuthal directions). 
We found that the effect of tongues tends to be stronger at the beginning of the evolution and weaker as the AR emerges. 
We also found a variety of evolutions that suggested, by comparison with the model, that emerging ARs have a wide set of twist profiles. 

Although the results obtained by \citet{Poisson16} provide constraints to theoretical and numerical models of FR emergence, they do not provide a method for removing the effects of the tongues from the intrinsic characteristics of emerging FRs. 
One of these intrinsic properties is the inclination of the emerging FR with respect to the solar equator, usually defined as the tilt angle. 
The distribution of tilt with solar latitude sets constraints on dynamo models.
In addition, there has been an important effort over the past few years to characterise the tilt of ARs with a diverse range of results \citep{Li12,McClintock13,McClintock14,Wang15,Tlatova18}. 
As we demonstrated in \citet{Poisson15b,Poisson16}, the presence of tongues has a non-negligible effect on the determination of the tilt.  
We consider that part of the disperse  (also taking into account the sometimes inconsistent results found by observational studies determining the tilt of ARs) can be affected by the lack of consideration for the effect of magnetic tongues in the measurements.

In this work, we introduce a method to compute the intrinsic tilt-angle of bipolar ARs using LOS magnetograms. The method aims to remove the effect of the magnetic tongues when computing the barycentres of the polarities and, hence, to obtain the AR tilt angle that better represents the intrinsic tilt.  
This method is defined in \sect{Method} following a summary of the  the analytical FR model used to test it. 
In \sect{CM}, we explore the parameter spaces of the model and of the method to evaluate the tilt correction and analyse the strengths and weaknesses of the method.  
Next, in \sect{Data}, we describe the four bipolar ARs used in \sect{Correction} to evaluate the method. 
These selected emerging ARs have strong magnetic tongues, which implies a strong deviation of the AR tilt deduced directly from the magnetograms.
Finally, in \sect{Conclusions}, we summarise and discuss our results.

\begin{figure}[!t]
\begin{center}
\includegraphics[height=.65\textwidth]{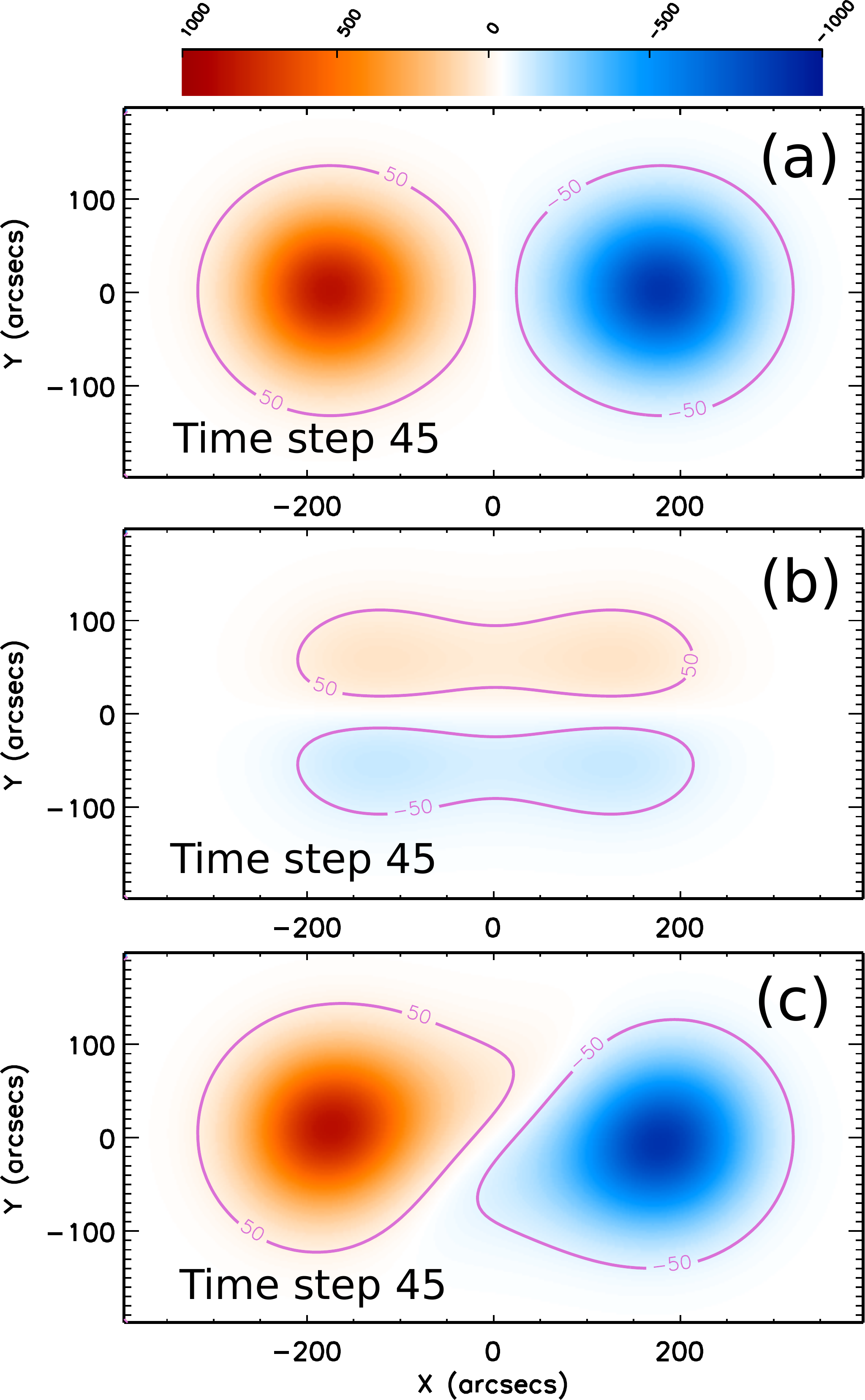}
\caption{Synthetic magnetograms of the axial (a) and the azimuthal (b) magnetic field components of a uniformly twisted torus model with ${\Nto } = 0.5$. 
(c) Total superposed magnetic field map.  
The red- and blue- shaded areas represent the positive  and the negative value of $\Bz$. 
The magenta contour in each map corresponds to $|\Bz| = 50$ G (the maximum axial field is set to 1000 G).
The associated movies are available online (fig1\_a.avi, fig1\_b.avi, and fig1\_c.avi).
}
\label{fig_AX-AZ}
\end{center} 
\end{figure}

\begin{figure*}
\begin{center}
\includegraphics[width=.9\textwidth,clip={150 0 150 0}]{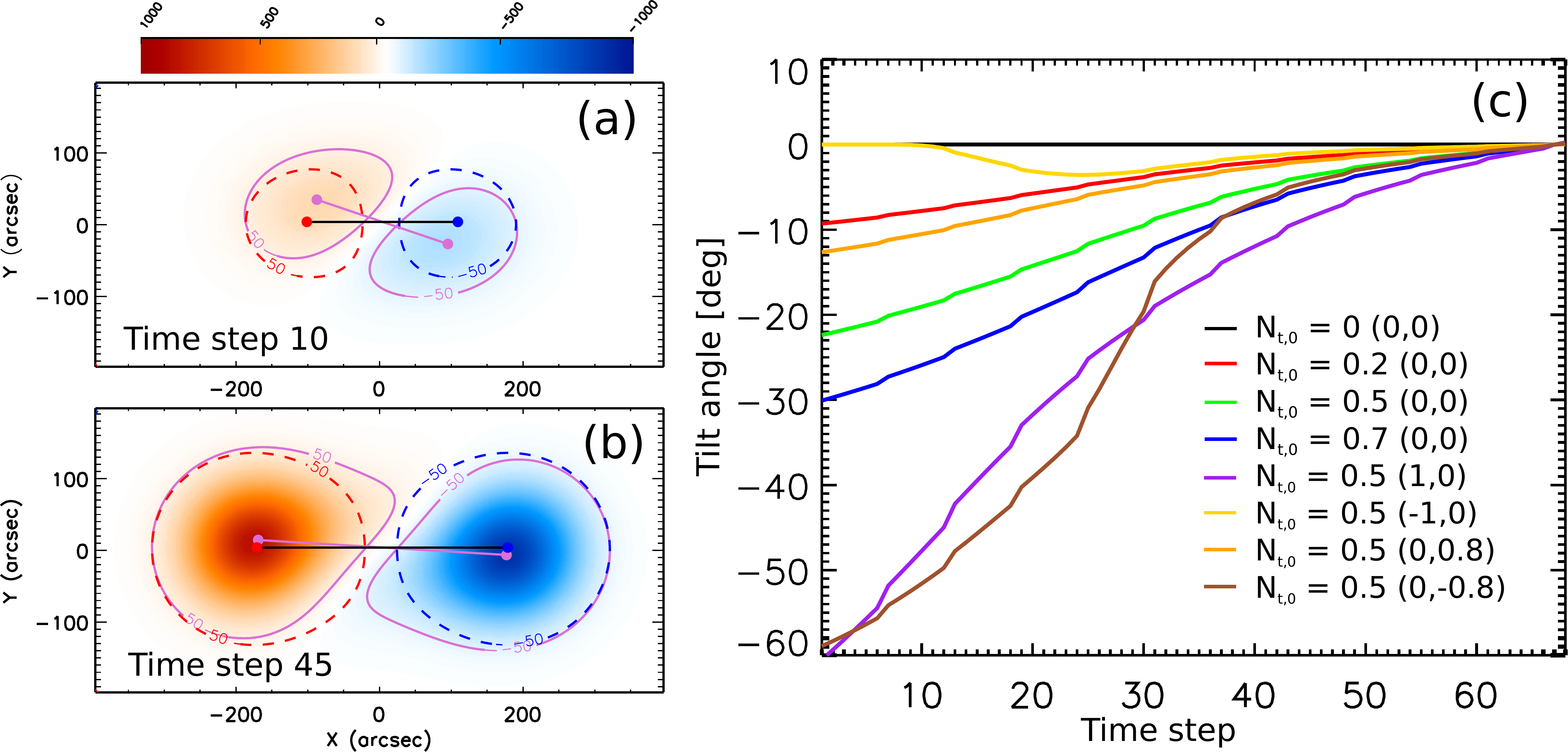}
\caption{(a)-(b) Synthetic magnetograms for the FR emergence model with ${\Nto } = 0.5$ ($h=0,g=0$) at two different time steps of the evolution. 
The red and blue dashed-contours correspond to the $|\Bz| = 50$ G isocontours of the axial field map (\fig{AX-AZ}a). 
The magnetic field strength on the FR axis is set to 1000 G.
The red and blue dots show the position of the magnetic barycentres of the axial map and the black segment in between indicates an intrinsic tilt $\phii = 0$.
The $|\Bz| = 50$ G isocontours of the total map (see also \fig{AX-AZ}c) are drawn with magenta continuous lines. 
The magenta line that joins the total map barycentres indicates an apparent tilt $\phia$.
(c) Evolution of $\phia$ for the uniform-twist models with ${\Nto } = 0, 0.2, 0.5, 0.7$ (black, red, green and blue continuous lines, respectively). 
The non-uniform twist models have a fixed value of ${\Nto } = 0.5$ and $(h,g)$ values indicated between brackets (violet, yellow, orange, and brown continuous lines, respectively).  
The associated movie is available online (fig2\_a.avi).
}
 \label{fig_model-tilts}
\end{center} 
\end{figure*}


\section{Flux rope model and the tilt correction method} 
\label{sect_Method}
  
In this section, we first summarise the main characteristics of the FR model defined in \citet{Luoni11}, then extended in \citet{Poisson16}, in order to set a framework to test the new method designed to remove the effect of magnetic tongues on the tilt of ARs.
  
\subsection{The simple flux emergence model} 
\label{sect_Model}

The simple FR model, developed in \citet{Luoni11}, provides a global description of the evolution of the photospheric magnetic field during the emergence of bipolar ARs. 
It consists of a toroidal FR with uniform twist (both along and across its axis). 
The sign and amount of magnetic twist is given by imposing a number of turns [$N_{\rm t}$] to the magnetic field lines. 
$N_{\rm t}$ corresponds to half of the emerging torus \citep[see Figure 2 in][]{Luoni11}.
The axial field component is supposed to have a Gaussian profile in the FR cross section (a distribution typically present in numerical simulation of FRs in the CZ). 
The upper half of the torus is set to progressively emerge without distortion. 
Therefore, this simple model does not take into account the deformations and reconnections occurring during emergence.  
 
The emergence of the FR at the photospheric level provides a series of synthetic magnetograms, which are analysed in exactly the same way as observed magnetograms (\sect{Correction}). 
The procedure consists in cutting the toroidal rope by successive horizontal planes {($z=$ constant, where $z$ is vertical coordinate)}, which play the role of the photosphere at successive times, and computing the magnetic field projection in the direction normal to these planes [$\Bz$]. 
{These synthetic magnetograms are the result of the superposition of the axial and azimuthal field components of the torus ($\Bp$ and $\Bt$, respectively) projected in the z-direction \citep[see an example in Figure 1 of][]{Poisson15a}.} 
Henceforth, each cut, that is, each magnetogram, is identified with a time step number in an analogous way to the observation time of ARs. 
For all the models that we use in this work, we set the number of cuts at 65. 
 
In order to identify the effect of tongues on the synthetic magnetograms, in \fig{AX-AZ}a--b we separate the axial and azimuthal field components for time step 45. 
The axial field component map (\fig{AX-AZ}a) has a mirror symmetry with respect to the y-direction which corresponds to the PIL direction.  
In this example, the FR is oriented in the x-direction and the tilt is null. 
When we add the azimuthal component map (\fig{AX-AZ}b), we obtain a magnetogram in which the polarities are elongated producing magnetic tongues (\fig{AX-AZ}c).
{This change is caused by the combination of two effects. In the top part of the positive polarity {($y>0$)}, both the azimuthal and axial field components contribute to a positive $\Bz$, thus enhancing it compared to the no-tongue case. 
In the bottom part of the positive polarity {($y<0$)}, the azimuthal component contributes to a negative $\Bz$ and thus partially cancels the positive $\Bz$-signal from the axial component. 
The reverse occurs in the negative polarity, with the enhancement of the negative polarity occurring in the bottom part ($y<0$) and the cancellation in the top ($y>0$) part.}
Both enhancement and cancellation of the field components, result in an asymmetric shape of 
each polarity, so in the presence of magnetic tongues. 

{ 
In \citet{Poisson16}, the aforementioned simple model was extended to interpret the evolution of the PIL and tilt angle in ARs with a large variety of twist profiles.
For half of a torus-shaped FR with a major radius [$R$], a minor radius [$a$], and a field strength on the axis [$B_{0}$], we define the azimuthal field in the natural coordinates of the torus $\{\rho, \phi, \theta \}$ as 
 
   \BE  \label{eq_Btheta}
   \Bt (\rho,\theta) = 2 \, \rho ~\Nt(\rho) \, B_{0} ~e^{-(\rho/a)^2} 
                       ~/ (R+\rho \cos \theta)  \,,
   \EE

\noindent where $\rho$ is the distance to the torus axis, {$\phi$ is the angle describing the position along the torus axis, and} $\theta$ is the rotation angle around this axis. The modifications we introduced include a non-uniform twist profile [$\Nt(\rho)$] along the radial direction of the FR cross-section that depends on a single non-dimensional parameter, called $h$. 
We define the non-uniform twist profile as
   \BE  \label{eq_Nt}
    \Nt(\rho) = \Nto ~ {\rm max} (1+h~(\rho/a)^2,0)  \,. 
   \EE
For $h=0$, the FR is uniformly twisted as the model presented in \citet{Luoni11}. A value of $h>0$ implies a twist more concentrated at the edge of the FR; as an example, if $h=1$ the twist at $\rho=a$ duplicates the value at the centre. Similarly, $h<0$ implies a twist decreasing from the axis to the FR border.

We introduce a simple $\theta$-dependence on the torus axial field, which produces a non-uniformity of the twist in the azimuthal direction:

   \BA 
   \Bp (\rho,\theta) &=& B_{0} ~ \asy (\theta) ~e^{-(\rho/a)^2}      
                         \,, \nonumber \\
   \mbox{\rm with}~~ \asy (\theta) &=& (1+g \cos \theta) 
                         \,, \label{eq_Bphi}
   \EA

\noindent where $g$ is the parameter controlling the non-axisymmetric level of $\Bp$. This parameter introduces a strong asymmetry between the top and the bottom parts of the FR cross-section. If $g>0,$ the twist is lower in the top part of the FR cross-section than in its bottom part and, conversely, if $g<0$.  
Both parameters, $h$ and $g$, vary within the interval [-1, 1] and strongly affect the evolution of the magnetic tongues \citep[see Appendix B in][]{Poisson16}.}

Figure \ref{fig_model-tilts}a-b shows two time steps in the FR emergence when ${\Nto } = 0.5$ and $h = g = 0$ (i.e. they correspond to the uniformly twisted FR). 
The intrinsic tilt angle [$\phii$] of this FR is $\phii = 0$; {it is} indicated with a black continuous line joining the barycentres of the axial field as the FR emerges. 
When the torus is crossing the photospheric plane in the first steps, the projection of the azimuthal field in the normal direction increases.
As a consequence, the flux of the tongues becomes stronger and shifts the position of the barycentres towards the elongated region of each polarity, producing a spurious increase of the tilt angle, that is, an apparent tilt angle [$\phia$]. 
The magenta dots show the position of the barycentres computed from $\Bz$ and the solid magenta line marks the tilt direction obtained from these barycentres.  
\fig{model-tilts}c illustrates the evolution of the tilt angle for the torus model with different values of {$\Nto $ and $(h,g)$.}  
In this panel, the green line corresponds to the model shown in the left panel. 
We conclude that the deviation of $\phia$ from $\phii$ could be significant. 

\begin{figure*}[!t]
\begin{center}
\includegraphics[width=.9\textwidth]{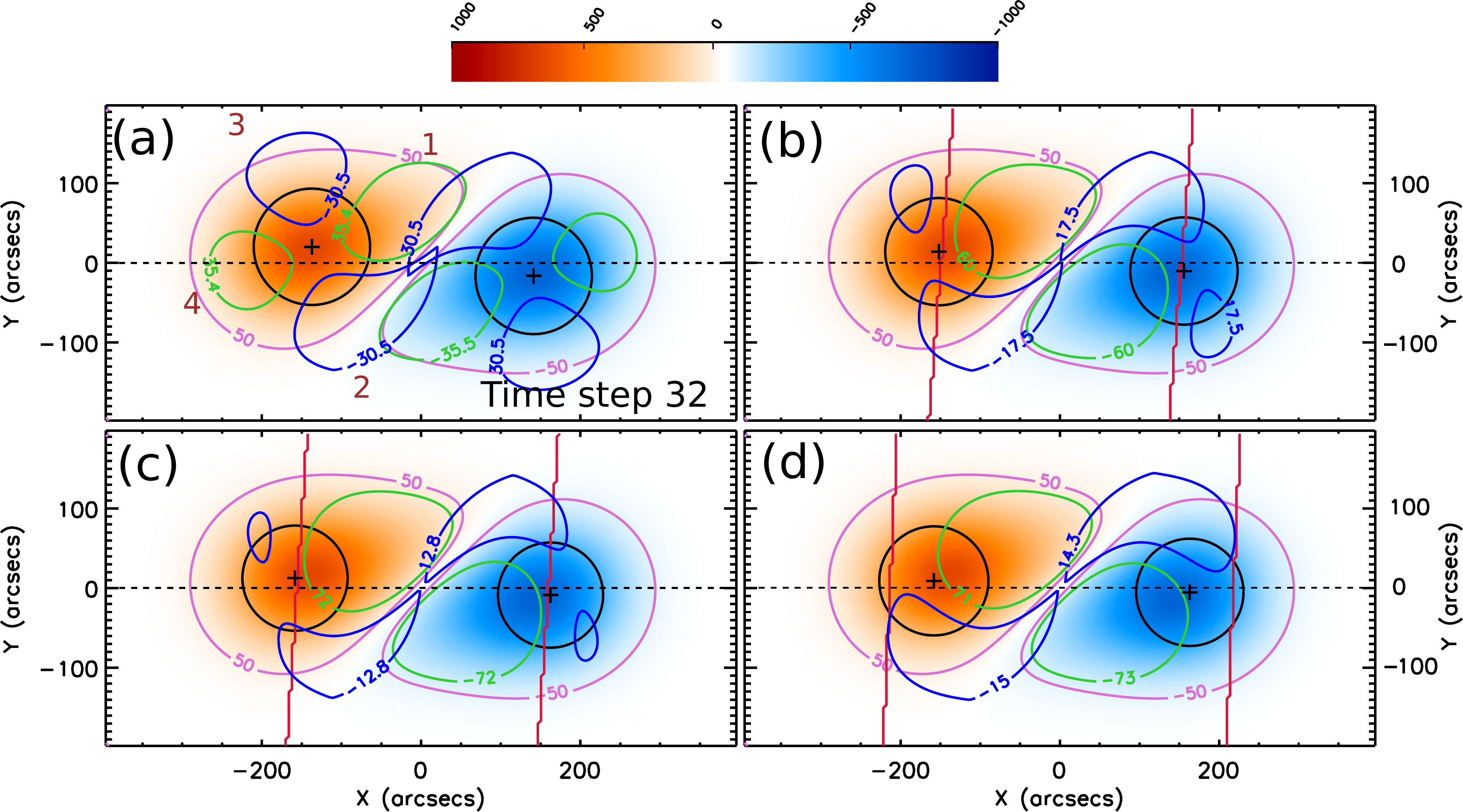}
\caption{(a) Synthetic $\Bz$ magnetogram for the torus model with uniform twist (${\Nto }=0.5$, $h =0$, $g = 0$) at time step $i = 32$. We use the same coloured code and isocontours levels for $\Bz$ than the ones used in \fig{AX-AZ}. 
The black circles over each polarity represent the contour level of $0.5 B_{\rm max}$ of the Gaussian function that best fits the magnetic field distribution. 
The position of the maximum value for each Gaussian is marked with a + symbol.
The green and blue isocontours indicate the areas where the field is larger and lower than the fitted Gaussian, respectively (by $50\%$ of the maximum difference between the Gaussian function values and the synthetic magnetogram).
 {The red lines, defined as discussed in \sect{GM},   delimit the area which is excluded from the computation of the Gaussian fit in the next iteration step.} 
(b) Same as (a) but one iteration step forward in the procedure ($n_{it} =1$). 
(c) Three iterations further ($n_{it}{=C_{it}} = 4$).
The associated movie is available online (fig3\_c.avi).
(d) Same as (c) but with the exclusion area extended one $\sigma$ (Gaussian width) towards the outer part of the {magnetic bipole} (see \sect{GM}).
}
 \label{fig_itera}
\end{center} 
\end{figure*}

\subsection{Core Field Fit Estimator}
\label{sect_GM}

The separation of the axial and the azimuthal field components from only a $\Bz$ magnetogram is an ill-posed problem.    
Since we cannot identify both components separately, we define the core and the tongue regions as the areas on the magnetogram that best represent the extension of the axial and azimuthal components, respectively, and we develop a method to approximately locate and constrain each of these regions on synthetic $\Bz$ magnetograms.  

The procedure aims to locate on each synthetic magnetogram the magnetic flux in the FR core region, which is the closest to the axial flux.
Henceforth, we refer to this method as the core field fit estimator (CoFFE).
The segment between the centres of the core flux region of each polarity is the one we use to estimate the FR tilt-angle [$\phic$]. One of the main advantages of using our FR model (\sect{Model}) to design this method is that we can test our procedure by comparing $\phi_i$ to $\phic$, which, in the best scenario, should be the same. 
 
We approximate the core field distribution assuming that it is symmetric with respect to the core centre position and with a decreasing profile towards the borders. 
{We use a Gaussian function for our procedure  because that is the profile of the axial field distribution in the simple FR model.} 
{ An axisymmetric Gaussian} function has five free parameters: the maximum field strength [$B_{\rm max}$], the magnetic baseline level [$B_{\rm 0}$], the position of the Gaussian centre [$x_{\rm max},y_{\rm max}$], and the mean dispersion around the centre, or simply the half width [$\sigma$].  
We fix $B_{\rm 0} = 0$, since the mean background field is null in our synthetic magnetograms.

The Gaussian function is fitted using an IDL routine based on the non-linear least-squares fitter {\sf MPFIT}. 
Since the effect of the azimuthal field component in emerging ARs typically decreases with time, as it is the case in the model, we first apply the fit to the last magnetogram of the series.
For the FR model, when the half torus is fully emerged, that is,{} at magnetogram 65, the Gaussian fit is exact due to the absence of the azimuthal component and the Gaussian distribution selected for the axial component.
Next we use the parameters of the fit at step $i$ as an initial guess to estimate the Gaussian for the magnetogram at step $i-1$.
This way of proceeding improves the stability of the method, assuming that consecutive magnetograms have similar magnetic flux distributions.

We now present how we proceeded and applied our method. 
Figure \ref{fig_itera}a shows the magnetogram for the model with uniform twist (${\Nto } = 0.5$) at time step 32, which is when the LOS projection of the azimuthal flux is maximum (\fig{TILTIT}a). 
The black circles correspond to {the contour levels drawn at half the maximum} of the Gaussian function fitted over each polarity.  
This fitting has been done using the information of the entire magnetogram.  
In this first step, we identify areas where the magnetic flux departs from the Gaussian distribution.  
We represent those areas with green and blue isocontours for the flux that is larger and lower by a given amount  than the fitted Gaussian, respectively.  We fix this amount to $50\%$ of the maximum difference between the Gaussian function values and the synthetic magnetogram.
These contours are labeled in \fig{itera}a on the positive polarity (symmetric contours are present on the negative polarity). 
The largest of these isocontours (labeled 1 and 2) represent the asymmetry generated by the presence of the azimuthal field.
The outer isocontours (3 and 4) are produced by the shift of the position of the Gaussian peak towards the tongues. 
These last isocontours should be very small; they should not even be present at all when we have a good approximation to the core field distribution.  

We use the information obtained from the first fitting (\fig{itera}a), which we call zero iteration, and redo the fit by masking those points where the asymmetry is the largest, that is,{} where the excess or lack of flux with respect to the Gaussian function is larger. 
To do this we select a region between the Gaussian centres as the one delimited by the red lines in \fig{itera}b and c. 
The red lines are defined with a linear function which crosses the core centre of the respective polarity and it is also orthogonal to the bipole axis defined by $\phic$ in the previous iteration.
The region defined between these lines is used to exclude those points from the map at the next iteration,
{{that is,} \fig{itera}b, which is obtained excluding a region that was defined at iteration zero but the lines shown in this panel enclose the region that is to be excluded in the next iteration}. In this way, we  consider the flux of the polarities with less contribution from the azimuthal field. 
The removal of the points is done using a weighting option; we assign a weight of 1 or 0 to the points that are outside or inside the limited region, respectively. 
We call this new computation {(see \fig{itera}b)} as the first iteration of the method. 
Once we get the new parameters of the Gaussian, a new limited region for the next iteration is defined (see \fig{itera}c). 

We repeat this procedure until we consider we have found the best fit to the core field distribution.
We set lower and upper limits to the number of iterations [$n_{it}$] we carry out for each magnetogram.  
We fix the lower limit to four iterations.  
We chose to stop the computation when $\phic$ varies by less than 1$\degree$ between two successive iterations.
We call $\Cit$ the minimum iteration in which this convergence criterion is fulfilled.
In summary, the method automatically iterates $\Cit$ times at time step $i$ and then uses the fitted parameters as a first guess for the parameters at time step $i-1$.

Figure \ref{fig_itera}c corresponds to the Gaussian fitting for $n_{it} = 4$.
One of the outer isocontours (4 in \fig{itera}a) has vanished but the other one is still present. 
Although the latter contour is smaller than in the first iteration (\fig{itera}b), the remaining flux below the Gaussian means that the points removed at each iteration are not enough to fully eliminate the effect of the tongues from the computed tilt angle. 
Therefore, we modify the region that is excluded using a single parameter $p$ that shifts away from the polarity centres each of the red lines by a quantity $p \,\sigma$ where $\sigma$ is the width of the fitted Gaussian.  If $p$ is positive, the region is to be extended.
Figure \ref{fig_itera}d shows the same magnetogram of panel c but with the Gaussian computed with $p=1$. 
The main difference between these two panels is in the size and location of the green and blue isocontours. 
Removing a larger area of the polarities means that we reduce the effect of the tongues when we fit the core field.
In \fig{itera}d, the increase of the green contours corresponds better to the location of the tongue flux. 
We can increase even more the exclusion region but at some point the amount of flux removed from the computation affects the goodness of the fit and the stability of the iterative procedure. 

{The standard error on the {position of the Gaussian centre, $x_{\rm max}$ and $y_{\rm max}$,} can be computed using the diagonal elements of the covariance matrix obtained with the non linear least-square fit.
But these errors {are too small and} do not reflect the goodness of the model to approximate the polarity core field.
In particular, the effect of the exclusion region is not well accounted for considering only the standard deviation of the fitting parameters.

Therefore, {we consider another type of error, more intrinsic to the CoFFE method.  We} compute $\phic$ using values of $p$ within the interval $[-1,1]$ to delimit the range of the tilt angle correction.
We consider $p=1$ as an upper limit due to its impact on the fitting procedure.
For $p>1$ the exclusion region removes more than $85\%$ of the core flux significantly degrading the goodness of the fit. {Conversely, for $p<-1$, more core flux is added to the fit 
but the tongue flux included also increases as $p$ becomes more negative.}}

\section{Correction of the tilt angle for a FR model}
\label{sect_CM}
        
In this section, we apply the CoFFE method (described in \sect{GM}) to isolate the core flux along the emergence of the modelled FR described in \sect{Model}.
We scan different twist profiles and we investigate how the results depend on the parameters of the CoFFE method.

\subsection{First tests} 
\label{sect_CM_First}

\begin{figure}[!t]
\begin{center}
\includegraphics[width=.4\textwidth]{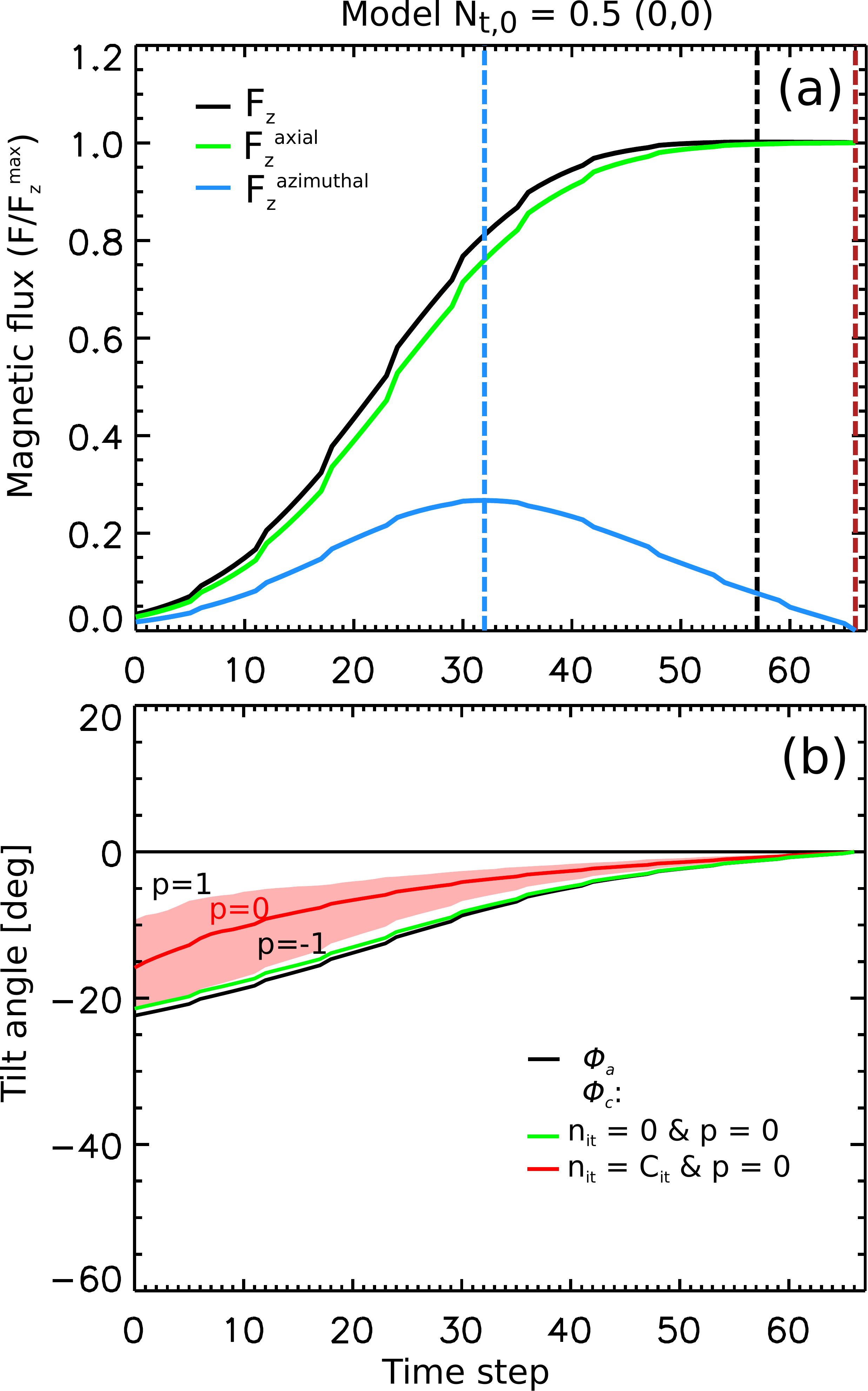}
\caption{(a) Evolution of the magnetic flux computed from the synthetic magnetograms for the FR model with uniform twist (${\Nto } = 0.5$, $h = g = 0$) during its emergence. 
We plot the magnetic flux [$\Fz$] (black line), the axial flux [$\Fzax$] (green line), and the azimuthal flux [$\Fzaz$] (blue line).  
All fluxes are computed for the z component of the fields.
All fluxes are normalised to the maximum axial FR flux. 
  (b) { Evolution of {the apparent tilt} $\phia$ computed from the magnetic barycentres (on the full magnetogram, black line) and $\phic$ using CoFFE  with $p=0$ for two $\nit$ {values} (see the inset).
The coloured-shaded {area surrounding the red line represents} the $\phic$ range obtained for different values of $p$ within the interval $[-1,1]$ after the convergence criterion of the tilt angle is achieved.} 
}
 \label{fig_TILTIT}
\end{center} 
\end{figure}

The evolution of the total magnetic flux [$\Fz$] of each AR polarity in the model with ${\Nto } = 0.5$ is shown in \fig{TILTIT}a (black line). 
The green and blue lines correspond, respectively, to the evolution of the FR axial and azimuthal flux contributions in the synthetic magnetograms. 
The difference between the black and green curves changes during the emergence. 
This difference corresponds to the effect of the tongues on $\Fz$.
The tongue effect peaks when the FR is about halfway emerged.

Even though the effect of the tongues becomes smaller as we go closer to the end of the emergence, the field enhancement produced by the tongues is enough to modify the maximum of $\Fz$. 
This maximum is reached before the half torus is fully emerged (see black-dashed vertical line in \fig{TILTIT}a) at step 57, while the effect of the azimuthal 
flux disappears at step 65 when the half the torus has emerged. 
We henceforth call this effect $\Bz$ enhancement. 
This implies that at the time of maximum flux we can still have strong magnetic tongues and, therefore, the time of maximum flux cannot be used to identify the end of the emergence (even if it seems to be a logical choice). 

\subsection{Convergence tests} 
\label{sect_CM_Convergence}

The corrected tilt angle [$\phic$] is determined by the acute angle between the $x-$axis and the line that joints the core centres with the CoFFE method, meaning its tongue effects have been removed, while {the apparent tilt} $\phia$ is the tilt deduced from the full magnetogram 
{using the barycentres}. 
Figure \ref{fig_TILTIT}b shows the computed $\phia$ and $\phic$ using the model with uniform twist {(${\Nto }=0.5,\, h=0,\, g=0$).} 
The green and red lines show the evolution of $\phic$ obtained by CoFFE with $p=0$ and for $n_{it}$ = 0 and $\Cit$, respectively. 
They correspond to the $B_z$ distributions shown in \fig{itera}a,c.
$\phic$ becomes closer to $\phii = 0$ as the number of iterations increases, which shows that the method converges well.
 
{ {To estimate an error for the tilt angle derived from the CoFFE method}, we compute $\phic$ for different values of $p$.
{For each model, we compute nine estimations of $\phic$ for equispaced values of $p$ within the interval $[-1,1]$.}
These values are displayed in \fig{TILTIT}b with a red-shaded area.
{As $p$ is more negative the tilt is closer} 
to the result obtained at the zero iteration {(green curve)}.
As it is expected the inclusion of more flux of the tongues ($p$ more negative) produces a smaller shift of the Gaussian centres towards the core {so less tilt correction}.
In contrast, {a larger positive $p$ value} increases the Gaussian centre displacement at each iteration achieving a more significant correction of the tilt angle.
Despite the broader range of $\phic$ obtained from time step 15 to 0, two aspects are noteworthy. 
First, the results with $p>0$ {(above the red curve with $p=0$)} are closer to $\phii$ along most of the FR emergence.
More specifically, we find that for $p=1$ the variation of $\phic$ reduces to $8\degree$ away from $\phii = 0$, compared to the variation of $\approx 20\degree$ for $\phia$ (black line).
This implies a decrease of the tongue effect by {$\approx 60\%$} when using the extended region $p=1$.
Secondly, the mean value of $\Cit$ computed along the emergence ranges between the minimum iteration number ($4$) and a mean of $5.15$ for all the different values of $p$; this means that the method has a fast convergence to a stable value of $\phic$  at each time step.}

\begin{figure}[!t]
\begin{center}
\includegraphics[width=.4\textwidth]{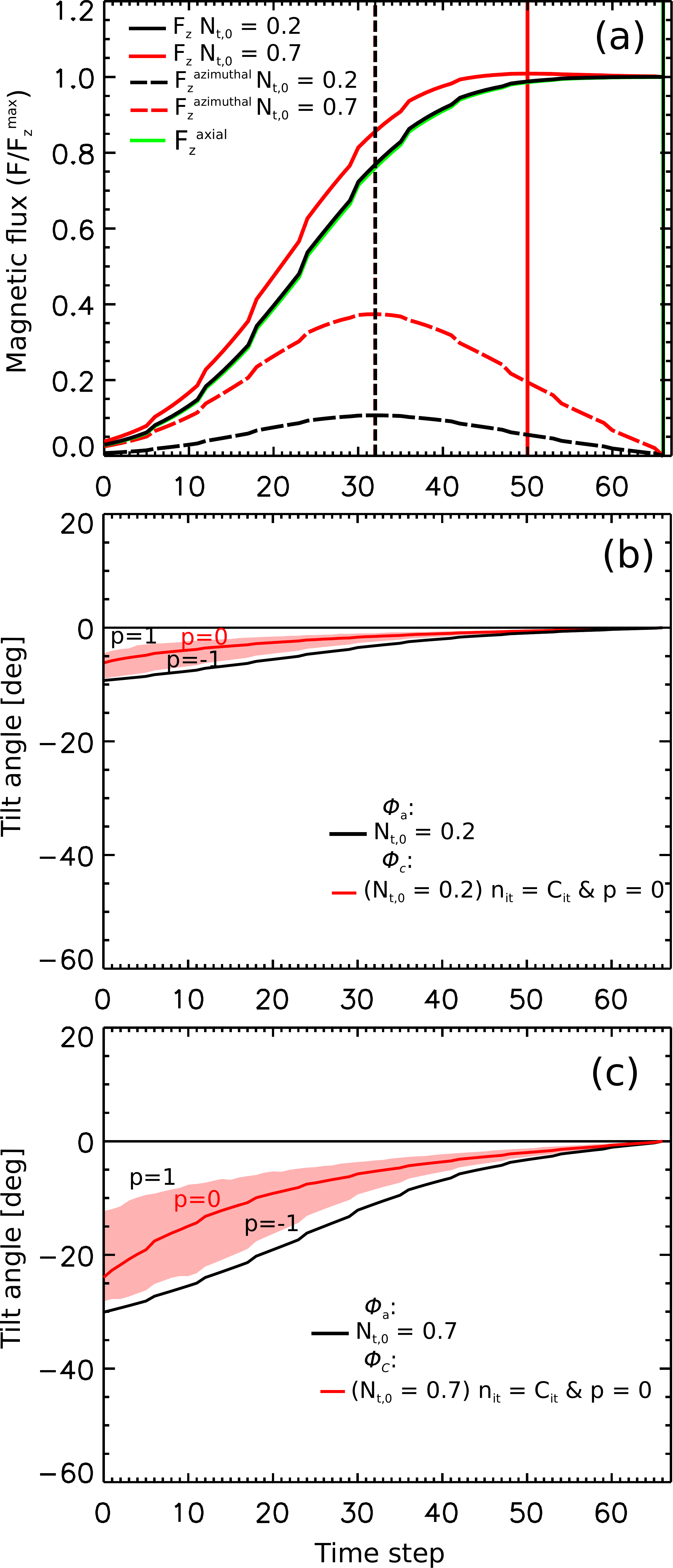}
\caption{ { Comparison of the magnetic flux and the tilt angle evolution for uniform twist models ($h=0$ and $g=0$) with ${\Nto }=0.2$ and ${\Nto }=0.7$.} 
(a) Evolution of the magnetic flux, $\Fz$ (continuous lines), the azimuthal flux, $\Fzaz$ (dashed lines), and the axial flux, $\Fzax$ (green-continuous line) for the models with ${\Nto } = 0.2$ (black) and ${\Nto } = 0.7$ (red).
The vertical continuous lines (with the same black and red colours) show respectively the time step in which each model reaches its maximum  $\Fz$ flux.
The black-dashed vertical line indicates the time step in which $\Fzaz$ is maximum.  
All fluxes are normalised to the maximum $\Fzax$.
{ (b)-(c) Evolution of the tilt angle for models with {(b) ${\Nto }=0.2$ and (c) ${\Nto }=0.7$.}
The black and red lines show the tilt angle estimations of $\phia$ and $\phic$ (see inset), respectively.
The red-shaded areas correspond to the values of $\phic$ computed for $p$ within the interval $[-1,1]$.}
The associated movies are available online (fig5\_b.avi and fig5\_c.avi).
}
 \label{fig_tilt-n}
\end{center} 
\end{figure}

\subsection{Effects of the twist amplitude} 
\label{sect_CM_Twist}

Next we selected different ${\Nto }$ values to test how the strength of the azimuthal field affects the stability of our method and the 
correction of the tilt angle. 
Figure \ref{fig_tilt-n} shows the magnetic flux and $\phic$ evolutions for models with low and high  uniform twist. 
In panel (a) we show the evolution of the magnetic flux (continuous lines) and the azimuthal flux (dashed lines) for both models, ${\Nto } = 0.2$ in black and ${\Nto }=0.7$ in red. 
The {maximum} azimuthal flux in the first case is around $10~\%$ of the total axial magnetic flux.  
Therefore, there is { no significant} $\Bz$ enhancement and the total maximum flux is reached at the end of the emergence at time step 65 (black line). 
On the contrary, the model with ${\Nto } = 0.7$ has a large enhancement of $\Bz$ that causes the shift of the maximum flux to time step 50 (red {vertical} line). 

In \fig{tilt-n}b--c we show the evolution of the apparent tilt $\phia$ for models with ${\Nto }=0.2$ and ${\Nto } = 0.7$ (black lines). 
Despite the reduced effect of the azimuthal flux for the case with ${\Nto } = 0.2$, there is a $10\degree$ variation of the apparent tilt angle during the emergence that 
departs from the FR intrinsic tilt, $\phii = 0$. 
The tilt variation for the ${\Nto } = 0.7$ model is around $30\degree$, that is $\approx 10\degree$ larger than the model with ${\Nto }=0.5$ (\fig{TILTIT}b). 

{ Next we use CoFFE with $\nit = \Cit$, {implying that convergence is achieved}, to compute a corrected tilt angle. 
The red lines in \fig{tilt-n}b--c show the evolution of $\phic$ computed with $p=0$ with ${\Nto }=0.2$ and ${\Nto } = 0.7$, respectively.
Using $p=0$ we reduce the difference between the corrected tilt $\phic$ and $\phii$ in approximately $45\%$ of the original difference derived from $\phia$ {(black line).
The percentage of correction is nearly} independent of the FR twist strength (see \figs{TILTIT}{tilt-n}).
The best correction of $\phic$ for both models is obtained with $p = 1$ {($\phic$ closer to 0)}. 
For the ${\Nto } = 0.2$ model, $\phic$ departs from the intrinsic tilt in less than $4\degree$ along the full FR emergence and even below $2\degree$ between time steps 20 to 50. 
{For} ${\Nto }=0.7$ the correction reduces the apparent rotation of the bipole to a $12\degree$ counter-clockwise rotation for the case $p=1$.
More generally, in all the models shown we are able to reduce the difference between $\phic$ and the intrinsic tilt of the bipole by {about} $60\%$ along the FR emergence. }

 {With even higher $p$ values, that is of{} $1.5$ and $2$ 
 (not shown here), and despite the large amount of flux removed from the fitting procedure, we are above an $80\%$ correction in all cases, reducing the difference between $\phic$ and $\phii$ below $10\degree$ along the full FR emergence for the model with ${\Nto } = 0.7$. However, the amount of flux removed from the procedure with these values of $p$ above $90~\%$ of the core flux makes the fit unstable, especially in real ARs where the core field does not follow in general a Gaussian profile.}

The FR model used has no deformation of its main axis (the FR axis is located in a plane). 
However, from the evolution of {the apparent tilt} $\phia$ we find a spurious rotation of the polarities produced by the magnetic tongues that can be wrongly related to the FR writhe, which corresponds to the deformation of the FR axis as a whole \citep{Lopez-Fuentes00}. 
All models with positive twist have an apparent rotation in a counter-clockwise sense due to the tongue retraction, suggesting a torus geometry with a positive writhe. 
Several studies have estimated the writhe of ARs from the evolution of the tilt angle along their emergence \citep{Lopez-Fuentes03,Liu14,Yang09}.
Our analysis suggests that the magnetic tongues have a strong impact on the correct estimation of the tilt in agreement with \citet{Lopez-Fuentes00}, who first show that the retraction of magnetic tongues could induced a fake rotation of an AR bipole, even reversing its direction of rotation during emergence. 
The correction achieved with CoFFE reduces this spurious rotation significantly, providing a more reliable estimation of the FR writhe.

\begin{figure*}[!t]
\begin{center}
\includegraphics[width=.8\textwidth]{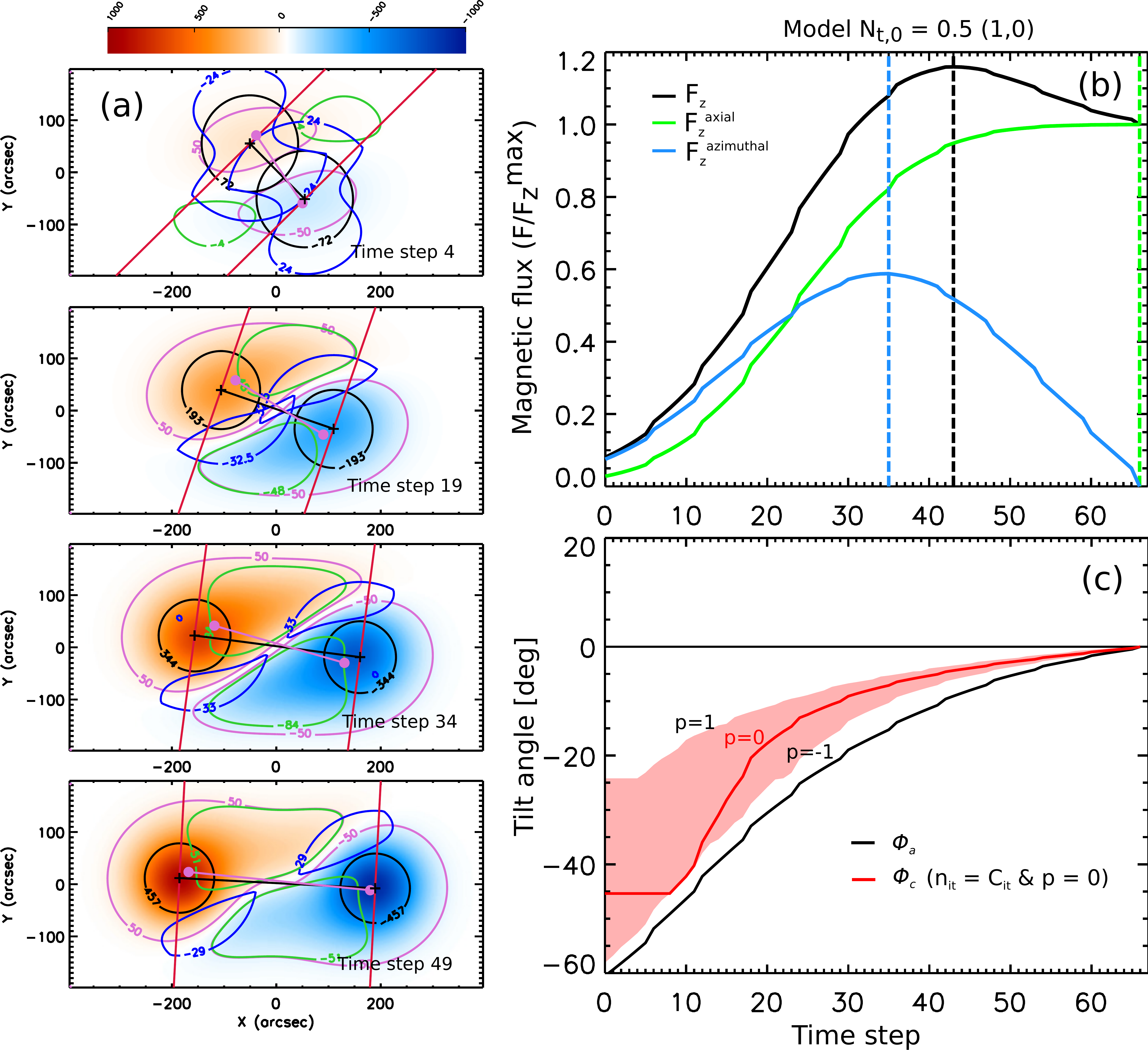}
\caption{(a) Synthetic $\Bz$ magnetograms for the model with a twist profile increasing with the radial coordinate {of the torus} (${\Nto } = 0.5$ and $h=1$, {defined in \eq{Nt}} at time steps $4$, $19$, $34$, and $49$). 
We use the same colour code and contours as in \fig{itera}. 
The segments join the polarities barycentres (magenta line) and the core centres (black line). 
(b) Computed $\Fz$ (black), $\Fzaz$ (blue), and $F_{z\rm }^{\rm axial}$ (green) along the FR emergence. 
Dashed-vertical lines indicate the time step of each maximum flux with their respective colour.
All fluxes are normalised to the maximum value of $F_{z\rm }^{\rm axial}$. 
(c) Evolution of the {apparent} tilt $\phia$ computed from the magnetic barycentres (black line) and of $\phic$ using CoFFE (see the inset). 
{The red-shaded area corresponds to the values of $\phic$ computed for $p$ within the interval $[-1,1]$ {after the convergence criterion of the tilt angle is achieved.}
The associated movie is available online (fig6\_a.avi).} 
}
 \label{fig_tilt-h}
\end{center} 
\end{figure*}

\subsection{Non-uniform twist} 
\label{sect_CM_Non_Uniform}

Next we study the model with non-uniform twist using the $h$ parameter as a measure of the radial profile of the twist (see \sect{Model}). 
We keep ${\Nto } = 0.5$ and we change $h$ using its extreme value of $1$ as reference. 
Figure \ref{fig_tilt-h}a shows the $\Bz$ map at different time steps for the model with $h = 1$ which corresponds to a flux-rope with an enhanced number of turns of $1$ at the periphery of the torus (in comparison to \fig{itera}).   
This increase of the azimuthal field is reflected in the strong elongation displayed by the tongues during all the emergence of the FR. 
In \fig{tilt-h}b the azimuthal flux (blue line) is larger than the axial flux (green line) till the first third of the emergence. 
The strong enhancement of the {azimuthal} magnetic flux shifts the maximum flux {(black dotted line)} to time step 43, far from the end of the FR emergence. 
In this case, the location of the maximum $\Fz$ is closer to the maximum azimuthal flux, which indicates that the tongues are strong at this time. 
This fact supports the statement mentioned above that the maximum flux is not a good reference to establish the centre location of the core field. 
At this time step the LOS projection of the azimuthal field masks most of the characteristic features of the axial field distribution. 

Figure \ref{fig_tilt-h}c shows that the apparent tilt angle, $\phia$ (black line), has a large variation of $60\degree$ due to the long and persistent magnetic tongues. 
 { $\phic$, derived from CoFFE with $p=0$ (red line), partially corrects the tilt angle achieving a variable correction ranging from $25\%$ at the first third of the emergence phase and up to $50\%$ towards the end of the emergence.  
Comparatively, the estimation for $p=1$ (red-shaded area upper boundary) produces a more stable computation of $\phic$ and provides about $55\%$ {correction of $\phic$ compared to $\phia$} along the FR emergence.}

This test case with $h=1$ represents an extreme instance where the azimuthal flux is larger than the axial flux during the first {third} of the emergence. 
The few ARs interpreted as formed by highly twisted FR develop other signatures on their flux distribution, apart from the magnetic tongues, that can be linked to the writhe of the main axis of the FR. 
In those cases, the magnetic tongues are difficult to interpret due to the development of kink instabilities that produce complex ARs with non simple bipolar magnetic field distribution \citep{Lopez-Fuentes03,Poisson13,Dalmasse13}. 
    
Next, the models with negative $h$ and positive $g$ have significantly reduced tongues { (see \fig{tilt-appendix}a-b in \app{extra})}, so the effect of the magnetic tongues tongues is comparable {or lower than} the already studies case with ${\Nto } = 0.2$.

Finally, a negative $g$ parameter produces a strong variation of {the apparent tilt} $\phia$ along the FR emergence { (see \fig{tilt-appendix}c in \app{extra})}.
Despite the large variation of $\phia$, this case {has results similar to} 
the case with strong magnetic tongues using the $h=1$ model {(\fig{tilt-h}).} 

\begin{figure*}[!t]
\begin{center}
\includegraphics[width=.9\textwidth]{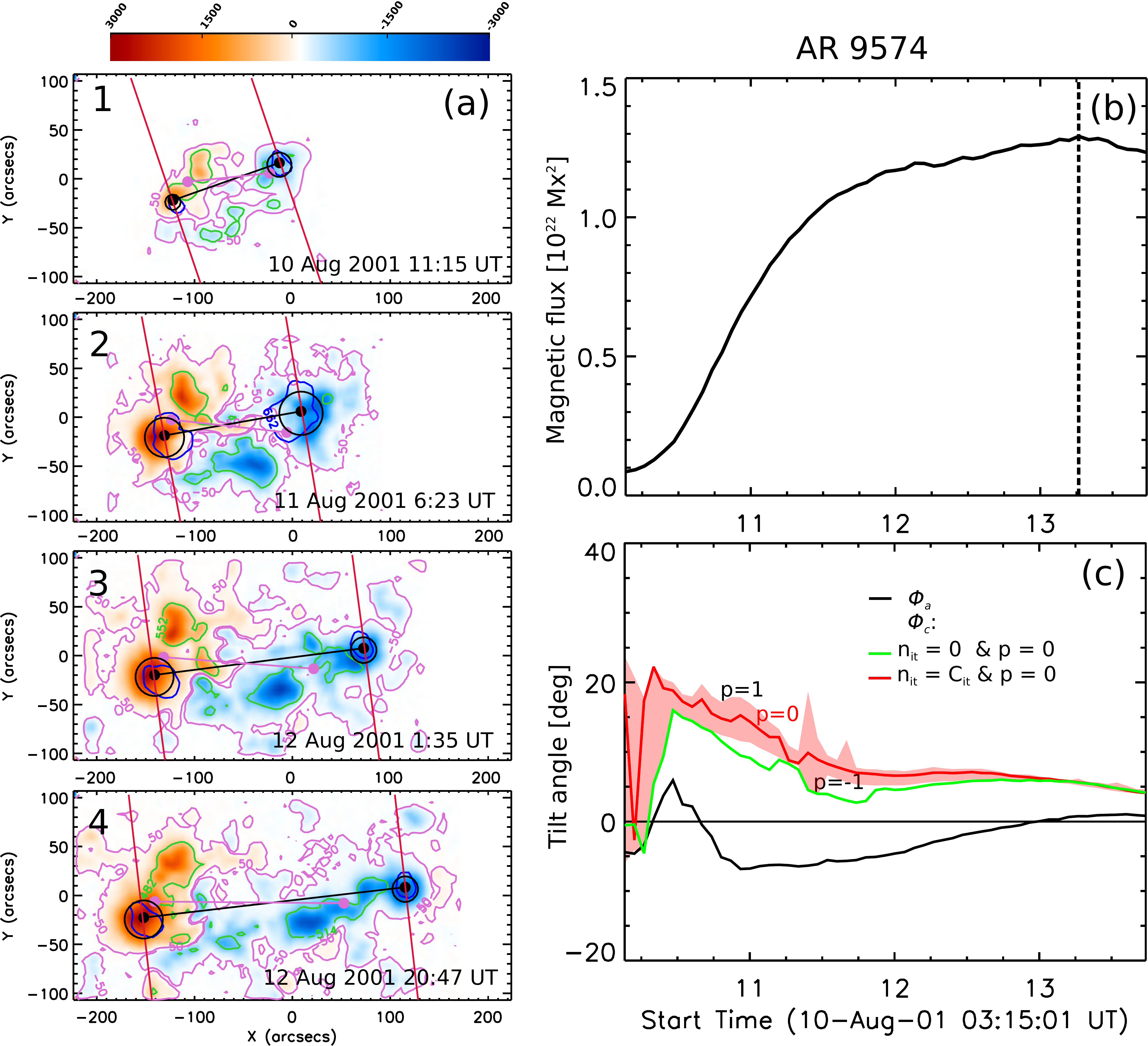}
\caption{(a) SOHO/MDI LOS magnetograms of AR 9574. 
The red- and blue- shaded areas represent the positive and the negative $\Bz$ magnetic field component. 
The magenta contour in each map corresponds to the field magnitude of {$\pm 50$ G}. 
The black circular contours are the half-width level of the Gaussian fit for each polarity using $p=0$ and $\nit = \Cit$. 
The green and blue isocontours indicate the areas where the field is larger and lower than the Gaussian, respectively, by $50\%$ of the maximum difference between 
the Gaussian and the local observed field. 
The black and the magenta segments show the tilt of the AR respectively obtained with the core flux centres (which define $\phic$) and the magnetic barycentres (which define {the apparent tilt} $\phia$). 
(b) Evolution of the AR magnetic flux computed from the magnetograms. {The vertical dashed line marks its maximum value.}  
(c) Evolution of $\phia$ derived from the magnetic barycentres (black line) and $\phic$ obtained with CoFFE (see inset). {The red-shaded area corresponds to the values of $\phic$ computed for $p$ within the interval $[-1,1]$ after convergence criterion of the tilt angle is achieved.}
The associated movie is available online (fig7\_a.avi).
}
 \label{fig_9574}
\end{center} 
\end{figure*}

\section{Data Used}
\label{sect_Data}

We applied CoFFE to LOS magnetograms of ARs obtained with the {\it Michelson Doppler Imager} \citep[MDI:][]{Scherrer95} on board the {\it Solar and Heliospheric Observatory} (SOHO).
The full-disc magnetograms with 96-minute cadence are obtained by averaging either one-minute and five-minute magnetograms. 
The different averages results in a flux density errors of 16 G and 9 G per pixel, respectively. 
These data provide 15 magnetograms per day with a size of 1024 $\times$ 1024 pixels and a spatial resolution of 1.98\arcsec. 

We use the same temporal and morphological criteria to select ARs as described in \citet{Poisson15b}. 
The sample consists of ARs with a low background flux, whose emergence phase is observed during their transit across the solar disc. 
We limit the latitude and the longitude of the emergence to {$\pm$ 30$^{\circ}$} from the solar equator or the central meridian (CM), respectively. 
This criterion aims to minimise foreshortening and projection effects when the AR is close to the solar limb \citep{Green03}. 
In order to describe the behaviour of our new method on observations, we select four bipolar ARs; these ARs are representative of regions with strong tongues. 
This selection is best to test the ability of CoFFE to remove the tilt deviation associated to magnetic tongues.
Furthermore, because CoFFE performs better when the tongues are less extended (lower tilt correction), we prefer not to exemplify such cases.

To obtain the set of magnetograms to which we applied CoFFE, we processed the full-disc magnetograms using standard Solar-Software tools.
We first transformed the magnetic-flux density measured in the LOS direction to the solar radial direction, neglecting the contribution of the components on the photospheric plane. 
This is a small correction as the selected ARs are close to the disc centre. 
The effects of this transformation were analysed by \citet{Green03}.  
Then we rotated the magnetograms to the date when the AR was located at the CM and we removed from the set any magnetogram with evidence of wrong pixels. 

Next we chose rectangular boxes of variable size encompassing the AR polarities during the emergence phase in order to minimise the contribution of the background magnetic flux \citep{Poisson16}. 
Movies for each AR were made to verify that the variable size box included all the magnetic flux of the AR at all times. 
All the AR parameters were computed considering only the pixels inside these rectangular boxes.

\section{Correction of the tilt angle for ARs}
 \label{sect_Correction} 
The application of CoFFE to the models in \sect{CM} has shown that $p$ is the most relevant parameter to obtain the best approximation to the intrinsic tilt of the bipole.
We have shown that a proper selection of $p$ can improve the computed tilt $\phic$ even in those cases where the azimuthal field is stronger than the axial field.
It is noteworthy however, to see that increasing the $p$ value also increases the uncertainty of the fitted Gaussian parameters {(because smaller portions of the magnetograms are used for the fits)}.
In this section, we present the $\phic$ results determined for four ARs and we test the dependence of the tilt correction on different values of $p$.

For each of the four selected ARs, we studied diverse conditions that help us to test the performance of CoFFE. 
We started with an AR which has a clear separation between the core and the tongue components, facilitating the core selection made with CoFFE. 
Then we analysed a more complicated case where the tongues are completely linked to the core distribution, as it happens for the FR model.
Increasing the tongue flux complexity, we studied the tilt correction in an AR where the tongues have a fragmented structure because of the emergence of a secondary bipole located at the centre of the AR. 
Finally, we tested CoFFE in an AR that has strong tongues all along the analysed time span {and a partially observed emergence limited by the longitudinal {criterion} of \sect{Data}.}

{As we have seen in \sect{CM}, we expect the core magnetic flux to be dominant at the end of the FR emergence.
However, while for observed ARs we may not be able to see the full emergence, we still use the last magnetogram of the set as the initial reference to start the computation.} 

\begin{figure*}[!t]
\begin{center}
\includegraphics[width=.9\textwidth]{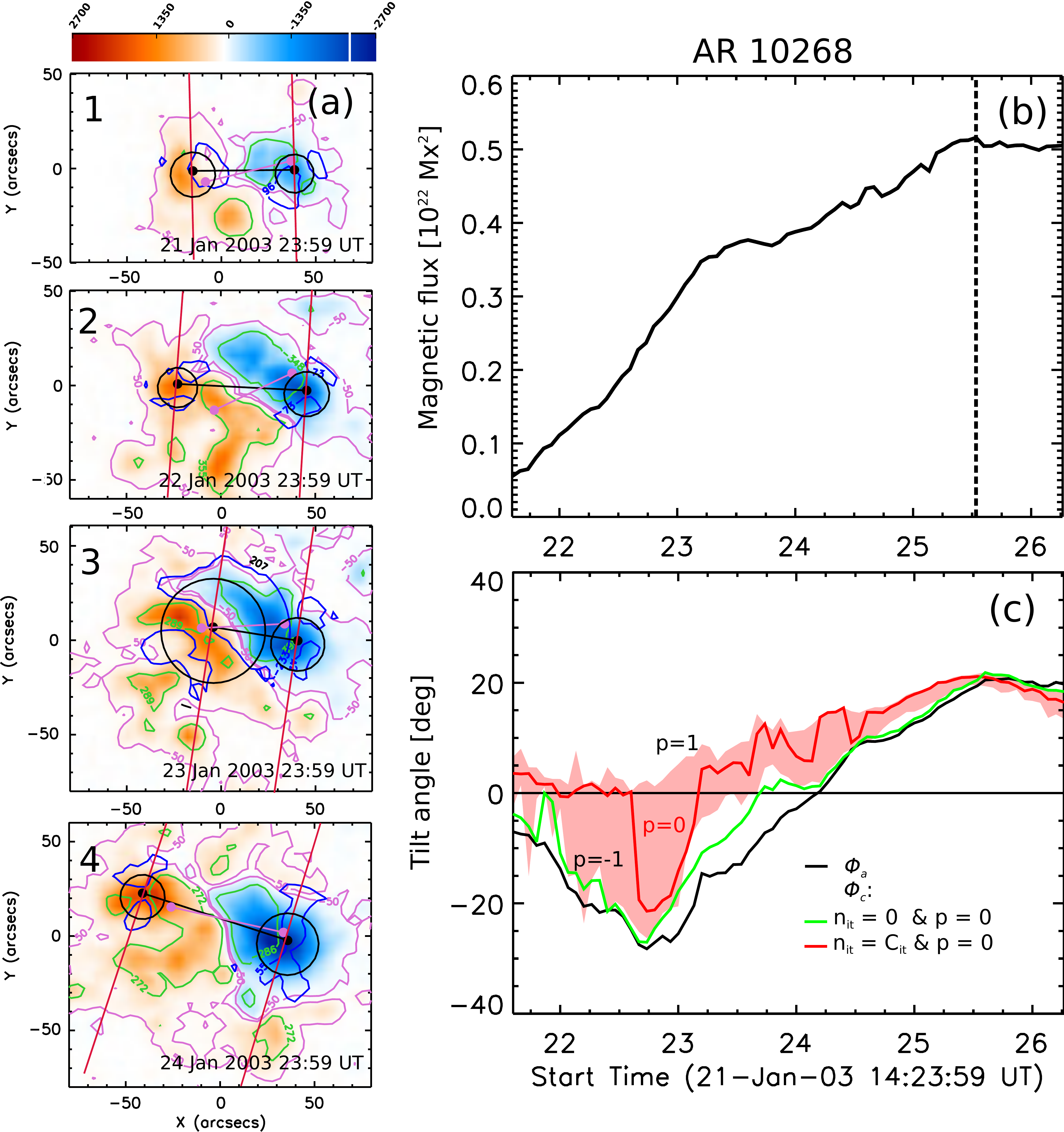}
\caption{ (a) SOHO/MDI magnetograms of AR 10268. 
The magnetograms have the same contours and colour convention as described in \fig{9574}a.
(b) Evolution of the AR magnetic flux computed from the magnetograms (black-solid line). 
The vertical-dashed line marks the time at which the AR 10268 reaches its maximum flux. 
(c) Evolution of $\phia$ derived from the magnetic barycentres and $\phic$ obtained with CoFFE (see inset). {The drawing conventions are the same as in \fig{9574}.}
The associated movie is available online (fig8\_a.avi).
}
 \label{fig_10268}
\end{center} 
\end{figure*}

\begin{figure*}[!t]
\begin{center}
\includegraphics[width=.9\textwidth]{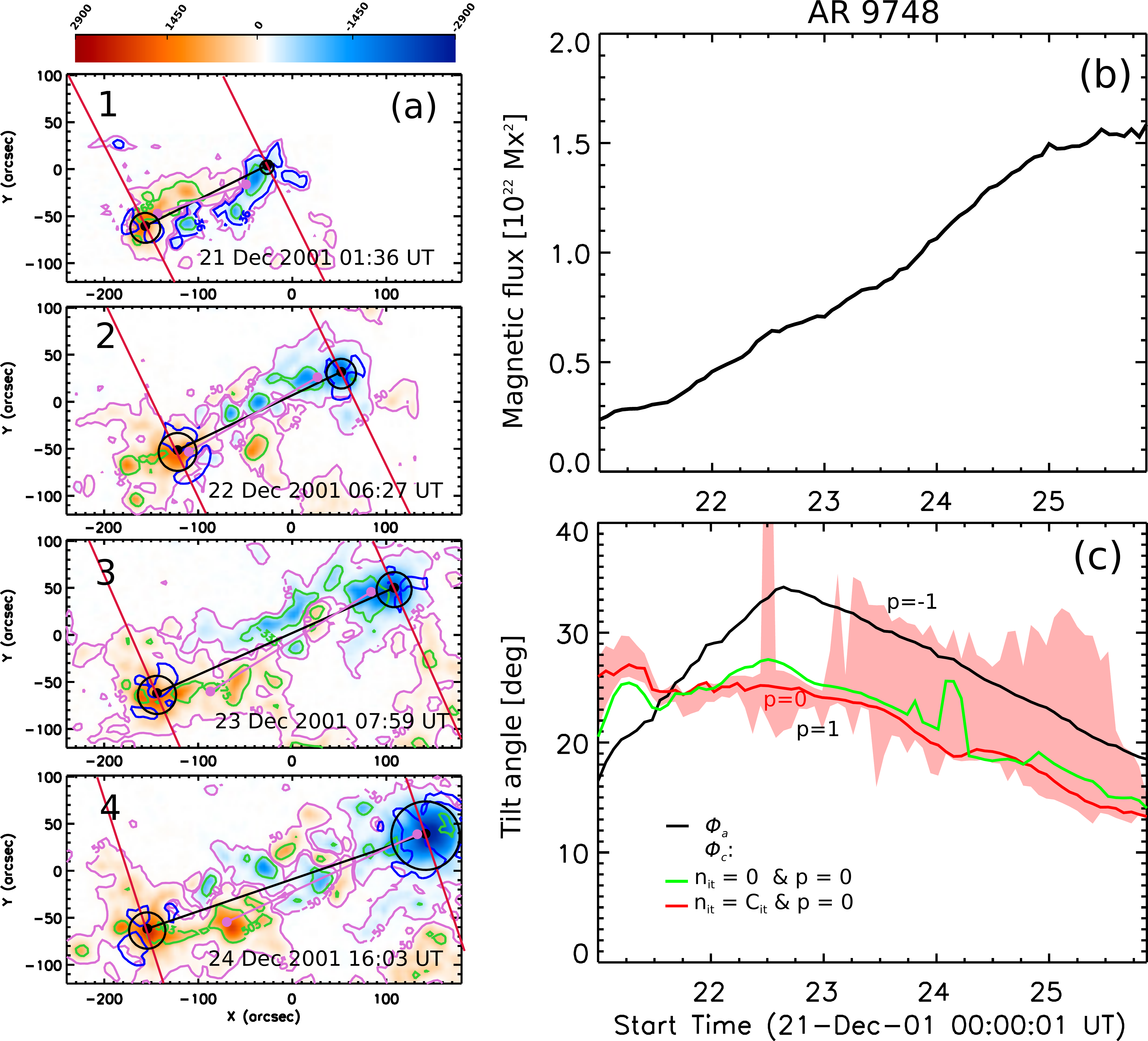}
\caption{ (a) SOHO/MDI magnetograms of AR 9748. 
The magnetograms have the same contours and colour convention as \fig{9574}a.
(b) Evolution of the AR magnetic flux computed from the magnetograms (black line). 
(c) Evolution of $\phia$ derived from the magnetic barycentres (black line) and $\phic$ obtained from the core flux centres with the same coloured pattern and input parameters as the ones used in \fig{9574}c. 
The associated movie is available online (fig9\_a.avi).
}
 \label{fig_9748}
\end{center} 
\end{figure*}

Despite the complexity of the observed ARs, magnetic tongues have a field distribution that is typically easy to recognise and locate in the LOS magnetograms by simple visual inspection.  
Conversely to the models of \sect{CM}, where the tongue and core fluxes have a continuous overlap, the tongue flux is typically more separated from the core flux in ARs.   {In the case of observed ARs no intrinsic tilt is available to test the CoFFE results, then, we  qualitatively} consider that a fit of the core is good when the field that is above the Gaussian, that is, the\ green contours in \fig{9574}a, \ref{fig_10268}a, \ref{fig_9748}a, and \ref{fig_9906}a approximately coincide with the observed magnetic tongues. 

It is worth noting that the flux of the core in ARs does not follow in general a Gaussian profile. 
Furthermore, this core flux profile varies from AR to AR; therefore, without having a clear flux profile for all ARs and in an effort to apply our method uniformly, we decided to use the profile that is better applied to the modelled FR for observed ARs. On the other hand, a Gaussian profile is the simplest and has the lower number of free parameters still keeping most of AR core characteristics (\ie\ a maximum of the flux and a width).


\subsection{Results for AR 9574} 
\label{sect_9574} 
    
AR 9574 emerged  between 10 and 13 August 2001. 
The time span for this AR encompasses 55 magnetograms limited according to the longitude criterion defined in \sect{Data}. 
This bipolar AR emerged in a low magnetic field environment and close to the equator (S$3$).  
Figure \ref{fig_9574}a shows a set of four MDI magnetograms corresponding to its emergence phase. 
The red- and blue-shaded areas on the magnetograms represent the strength of the outward and inward components of the radial magnetic field. 
We refer to this field as $\Bz$ as it is equivalent to the field on the synthetic magnetograms. 
The magenta isocontours correspond to $|\Bz| = 50$ G. 
Elongated polarities associated to strong magnetic tongues are present all along the AR emergence. 
Despite the asymmetric elongation between the leading and the following polarity, we recognise a magnetic tongue pattern that corresponds to a FR with positive twist. 

CoFFE provides the approximate location and extension of the core field on each polarity.  
The circular black contours in \fig{9574}a correspond to $50\%$ of the fitted parameter $B_{\rm max}$ on each polarity. 
The location of $B_{\rm max}$ for each polarity is marked with a black dot. 
The core field distribution {is} obtained using $p=0$ and $\nit = \Cit$. 
For these parameters, we obtain the best tracking of the core field centre back to the first flux emergence of AR 9574 (see ``fig7\_a.avi'' in the supplementary material).

Green and blue contours show the regions where $\Bz$ is significantly above and below the Gaussian values, respectively. 
The values of those contours are expressed in Gauss over the drawn lines and set on each magnetogram as half the maximum difference between $\Bz$ and the fitted Gaussian. 
The green contours outline the spatial extension of the magnetic tongues (\sect{CM_First}). 
The blue contours are dominant on the core regions indicating that the Gaussian profile is not completely correct to approximate the field closer to the core centre.
As a consequence of this discrepancy, we notice that the observed field on the centre of the polarity is less concentrated than the Gaussian function determined with CoFFE.

CoFFE excludes a region around the central PIL of the AR from the tilt computation (\sect{GM}). 
The red lines show the limits of the exclusion region in \fig{9574}a with $p=0$. 
Finally, we show the lines that join the core field centres (black segment), and the magnetic barycentres of the full magnetogram (magenta segment) to compare $\phic$ and $\phia$ tilts. 

The emergence near the central meridian passage allows us to analyse the evolution starting from the first flux emergence, that is, when the AR flux is below 10\% of the maximum flux observed. 
A fast flux emergence rate is present in the first half of the evolution in association to the early development of strong magnetic tongues (\fig{9574}b). 
The last half of the emergence has a lower flux emergence rate similar to what we obtained for different models in \sect{CM}. 
The latter evolution corresponds to the progressive reduction of the tongue magnetic flux. 

Figure \ref{fig_9574}c shows the evolution of the tilt angle computed along the emergence of AR 9574.
We define a positive tilt angle when the leading polarity is closer to the equator than the following one, consistently with the Joy's law. 
The method with no iterations ($\nit = 0$) and $p=0$ (green line) performs similarly as the other cases with higher $\nit$ and $p$ parameters, except at the beginning of the emergence (see below). 
This implies that CoFFE removes the tongues in a similar way independently of the parameter values so that the core flux is properly identified. 
This is the case because the spatial separation between the tongues and the core distributions is more pronounced than for the modelled cases in \sect{CM}, as seen by comparing \fig{9574}a with \fig{itera}. 
\citet{Archontis10} described a fragmented configuration of the tongues due to the convergence of vertical plasma flows on the photosphere producing tongue regions where the field is compressed.
This produces a discrete structuring of the tongues that makes them more distinguishable than in the FR model.

Both cases, with $\nit = 0$ or $\nit= \Cit$, and $p=0$ have a consistent and clear evolution along most of the AR emergence, but they cannot track back the core centre when $\Fz$ is below {about} 10\% of its maximum. 
That short time span can only be corrected with CoFFE using $\nit = \Cit$ and the extended region of $p=1$ { (upper limit of the red-shaded area).}  

The tilt $\phic$ departs significantly from {the apparent tilt} $\phia$ for all the time span of the AR emergence. 
The difference between $\phic$ and $\phia$ ranges from $5\degree$ up to $25\degree$ (\fig{9574}c), which is above the uncertainties estimated for CoFFE {(red-shaded area)}. 
We also find that there is an opposite tilt of the bipole determined with CoFFE ($\phic$) and the magnetic barycentres ($\phia$). 
The mean $\phia$ is $\approx-2\degree$ while the mean for $\phic$ is approximately $10\degree$ (for $p=1$). 
The estimated values of $\phic$ correspond to an AR located at the southern hemisphere accordingly to Joy's law and indeed AR 9574 is in the southern hemisphere. 

Moreover, there is an opposite tendency for the bipole rotation determined with $\phic$ and $\phia$, suggesting different signs of FR writhe for each estimation. 
The rotation computed with $\phic$ is around $15\degree$ in the clockwise direction (negative writhe), while the evolution of $\phia$ indicates a counter-clockwise rotation of $5\degree$ (positive writhe). 
In this case the magnetic tongues affect the determination of the tilt angle changing both the intrinsic tilt and the rotation direction of the bipole.


\subsection{Results for AR 10268} 
\label{sect_10268}

Figure \ref{fig_10268}a shows four magnetograms of the emergence phase of AR 10268, each of them labelled with a number that indicates their temporal order.   
AR 10268 emerged in the north hemisphere (N12) around midday of 21 January 2003. 
The 5-day long evolution of the AR is only limited by the longitudinal criterion (see \sect{Data}), restricting the number of analysed magnetograms to 71. 
This AR has a clear bipolar configuration and magnetic tongues that are more interwoven with the main core field than in AR 9574.  
There is a small secondary bipole emerging northward of the leading polarity (magnetograms 3 and 4 in \fig{10268}a), but this has only a weak impact on {the apparent tilt} $\phia$, as the magnetic tongues, with their larger fluxes, are the main cause of the shift of the magnetic barycentres towards the AR centre (see also ``fig8\_a.avi'' in the additional material). 

As in AR 9574, AR 10268 has two different flux emergence rates (\fig{10268}b). 
The first part is associated with the fast development of tongues and the consequent enhancement produced on the observed field $\Bz$. 
After January 23, around the mid-emergence phase, there is a decrease in the flux emergence rate.

The evolution of {the apparent tilt} $\phia$ (\ie\ without tongue removal) shows that AR 10268 {rapidly evolves towards} a highly tilted bipole which opposes to Joy's law prediction (\fig{10268}c). 
The largest departure from the E-W direction reaches up to $-25\degree$. 
Towards the end of the emergence $\phia$ tends to a more regular tilt value, in agreement with Joy's law, $\approx 20\degree$.
However, a different tilt evolution is obtained after removing the tongues with CoFFE method as $\phic\approx 0\degree$ initially and it grows progressively to $\phic\approx 20\degree$ (\fig{10268}c).
For {$p$ within $[0,1]$} we get a consistent evolution of $\phic$ and a large departure from $\phia$ during more than the first half of the emergence phase.  
Later on, when the tongues retract (\fig{10268}a), $\phic$ values are consistent with $\phia$ (\fig{10268}c).
The difference between CoFFE and the barycentres estimation ranges from around $25\degree$ to $1\degree$ towards the end of the observed emergence. 
Finally, the mean number of iterations along the AR emergence is below $6$, which implies a fast convergence of the procedure to a stable and consistent $\phic$.  
 
The rotation of AR 10268, derived from $\phia$ has a sudden change during the first half of the emergence (see black line in \fig{10268}c).  
Indeed, $\phia$ measurements imply a spurious 20$\degree$ counter-clockwise rotation observed until 23 January and then a large clockwise rotation of $\approx 45\degree$ towards the end of the emergence. 
Using the $\phic$ estimation for $\nit > 0$ {and $p > 0,$} we are able to completely remove the initial spurious rotation and we get a consistent clockwise rotation of $\approx 20 \degree$ along the full emergence. 
We achieve a better approximation to the intrinsic rotation of the bipole, which is consistent both with the Joy's law and with the emergence of a FR with a negative writhe. 
In this case the signs of the twist and the writhe coincide. 


\begin{figure*}[!t]
\begin{center}
\includegraphics[width=.9\textwidth]{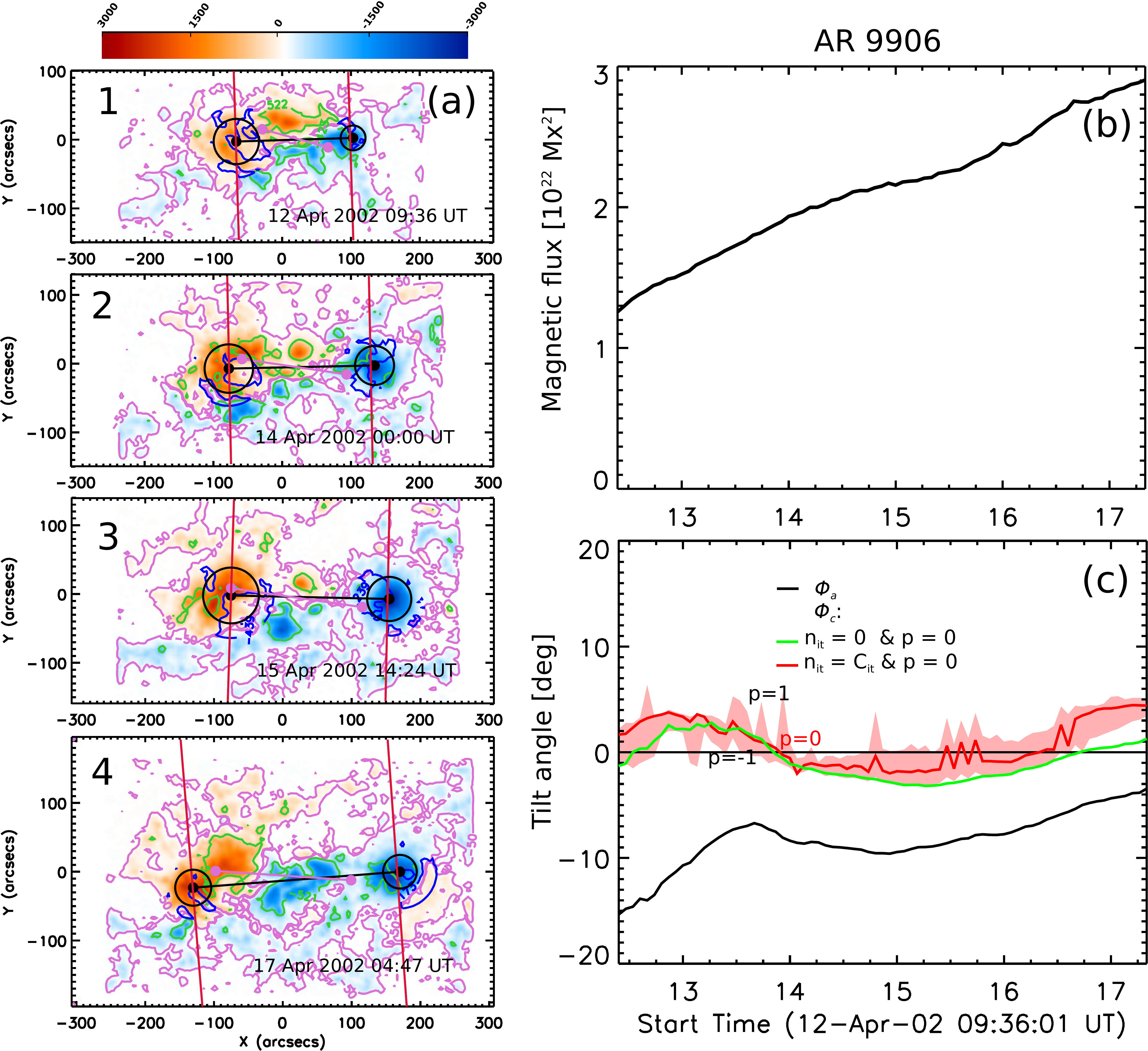}
\caption{ (a) SOHO/MDI magnetograms of AR 9906. 
The magnetograms have the same contours and colour convention as \fig{9574}a.
(b) Evolution of the AR magnetic flux computed from the magnetograms (black line). 
(c) Evolution of {the apparent tilt} $\phia$ derived from the magnetic barycentres (black line) and $\phic$ obtained from the core flux centres with the same coloured-pattern and input parameters as the ones used in \fig{9574}c.
The associated movie is available online (fig10\_a.avi).
}
 \label{fig_9906}
\end{center} 
\end{figure*}

\subsection{Results for AR 9748} 
\label{sect_9748}

AR 9748 emerged in the southern hemisphere (S11) and close to the east limb on 21 December 2001. 
We analyse its five-day long evolution across the solar disc which corresponds to 72 LOS magnetograms. 
The last magnetogram used corresponds to a longitudinal position close to W$30\degree$, so the evolution is truncated before reaching the end of the emergence. 
Figure \ref{fig_9748}a shows four magnetograms illustrating the evolution of AR 9748 magnetic flux distribution.
From the initial stage of the emergence a bipolar distribution is present with a large tilt with respect to the equatorial direction. 
The elongated magnetic tongues pattern is more complex than in previous examples as it is formed of fragmented polarities {that change of location}. 
For most of the AR evolution we recognise a tongue pattern which indicates that the AR is produced by the emergence of a FR with negative twist {(\eg\ \fig{9748}a on 23 December), while the initial tongue pattern at the beginning of the emergence rather indicates a FR with positive twist (\eg\ \fig{9748}a on 21 December)}. 

The core fit shown in \fig{9748}a was obtained with $p=0$ and {the number of iterations for convergence ($\nit = \Cit$)}.
The magnetic tongues mostly coincide with the green contours, although they seem to be mixed with other field elements, like a secondary bipole and/or magnetic field remnants present between both core field centres (see ``fig9\_a.avi'' in the additional material). 
Despite the complex field distribution and strong tongues observed in the last magnetogram, the method successfully locates the main core field contribution and it effectively tracks it back along the full observed evolution of the AR (\fig{9748}c).

AR 9748 has nearly a constant magnetic flux emergence rate along most of the observed evolution.
Only near the end there is a small decrease of this rate (\fig{9748}b). 
Comparing this flux evolution with the FR model examples we infer that the studied interval corresponds to about the first half of the emergence phase shown in Figs. \ref{fig_TILTIT}a and \ref{fig_tilt-n}a.
The time at which the flux emergence rate changes coincides with the maximum extension of the magnetic tongues (see ``fig9\_a.avi'' in the additional material).

\begin{table*}
 \caption{List of the analysed ARs and their computed parameters using the magnetic barycentres 
 and CoFFE. Column one and two show the AR NOAA number and the twist sign [$sign(T)$] derived from the magnetic tongues. 
 Columns three to seven list the values of the mean $\phia$ and $\phic$, the mean difference between both tilt estimations [$\overline{|\phia -\phic|}$], and the variation of the tilt angle 
[$\Delta \phia$] and [$\Delta \phic$] per day; all these values are expressed in degrees.
Finally, the right column shows the mean number of iterations [$\overline{n}_{\rm it}$] required for the computations. 
{The errors reported in columns three to five correspond to the standard deviation of the temporal averages.}
The values correspond to $p=0$. 
}
\begin{tabularx}{\textwidth}{@{}r*{8}{C}c@{}}
\toprule
AR &$sign(T)$& $\overline{\phi}_a$  & $\overline{\phic}$ & $\overline{|\phia -\phic|}$    & $\Delta \phia$ / day & $\Delta \phic$ / day   & $\overline{n}_{\rm it}$ \\ 
\midrule
9574 &+& -2$\degree \pm 3\degree$ & 10$\degree \pm 6\degree$ & $13\degree \pm 6\degree$ & 0.2$\degree$ & -3.9$\degree$ & 5.4 \\ 
9748 &-& 26$\degree \pm 4\degree$ & 20$\degree \pm 5\degree$ & $ 5\degree \pm 2\degree$ &-0.3$\degree$ & -2.7$\degree$ & 5.3 \\ 
9906 &+&-10$\degree \pm 4\degree$ &  2$\degree \pm 2\degree$ & $ 9\degree \pm 3\degree$ & 2.2$\degree$ &  0.9$\degree$ & 5.5 \\ 
10268&-& -1$\degree \pm 16\degree$ & 10$\degree \pm 9\degree$ & $13\degree \pm 9\degree$ &-6.0$  \degree$ & -2.6$\degree$ & 5.5 \\ 
\bottomrule
\end{tabularx}
\label{tableta}

\end{table*}

Figure \ref{fig_9748}c shows the tilt $\phia$ computed from the magnetic barycentres (black line) and the estimations of $\phic$ with the same set of parameters as in the previous {AR} example. 
For {$p$ in $[0,1]$,} 
we obtain similar results while there is a clear difference between $\phia$ and $\phic$ that ranges between $4\degree$ and $7\degree$ and is larger than the estimated uncertainties.
As seen in all the previous examples, the method performs efficiently requiring a low number of iterations to achieve a stable $\phic$ value. 
The mean number of iterations along the AR emergence is below $5.5$ in the cases where $p=0$ and $p=1$. 

$\phia$ measurements imply a strong apparent counter-clockwise rotation of the polarities of around $14\degree$ due to the presence of the magnetic tongues, which can be observed along the first day of the emergence. 
This rotation is removed when we compute $\phic$ and instead we find a constant clockwise rotation of the bipole of $\approx13\degree$ (\fig{9748}c). 
The inversion of the rotation sense observed with $\phia$ during the first analysed day of AR 9748 evolution, cannot be explained as the emergence of a single coherent FR.
This kind of change would imply an unrealistic scenario where the FR have mixed writhe along its main axis. 
The results achieved with CoFFE are consistent with a single emerging FR.

\subsection{Results for AR 9906} 
\label{sect_9906}

Finally, we tested CoFFE with the partial emergence of AR 9906.
During its solar disc transit, AR 9906 (S16) presented strong magnetic tongues with a pattern associated to a FR with positive twist.  
In this case, we were able to study only part of the emergence as we could not include the initial flux emergence and the maximum flux was reached far beyond the limit imposed by the longitude criterion.
The temporal interval covers five days between 12 and 17 April 2002.

Figure \ref{fig_9906}a shows selected magnetograms of the emergence phase of AR 9906. 
The full set of 75 magnetograms is shown in the movie ``fig10\_a.avi'', which is included in the additional material. 
We find that CoFFE provides a good approximation of the core field distribution for $p=0$ and {achieving convergence ($\nit=\Cit$)} at all times studied. 
{Both the $\Bz$ field regions above and below the fitted Gaussian, defined as in the previous examples, can be well identified by the green and the blue contours, respectively.}
Despite their small extensions, the blue contours in \fig{9906}a (panels 1, 2, and 3) show the regions in which the core field symmetry is affected by the tongue field of the opposite polarity.
The locations of these regions are similar to those observed for the FR model in \figs{itera}{tilt-h}, but as it was mentioned previously with regard to AR 9574, there is a discrepancy between the Gaussian profile used and the field distribution around the core centre. {However, the green contours are well delimiting the magnetic tongues.} 
Hence, all the field contributions are well isolated by the method. 
This example has the field distribution that most resembles the model used in \sect{CM} (see \fig{tilt-h}a).

This AR has a slow emergence rate that allows us to have a good temporal resolution of the first half of the emergence. 
Figure \ref{fig_9906}b shows that the magnetic flux increases with a constant rate.
Comparing this case with the other ARs, the magnetic flux evolution suggests that the AR does not reach the time corresponding to half FR emergence (see \sect{Model}). 
Interestingly, this example shows how CoFFE responds in a case in which the tongues flux is as strong as the core flux at all times. 

Figure \ref{fig_9906}c shows the $\phia$ evolution (black line) and the $\phic$ estimations using CoFFE with the same set of parameters and colour convention as for the previous ARs analysis.  
The difference between both tilt angle estimations ranges between $16\degree$ to $5\degree$, being around $8\degree$ at the last analysed magnetogram.
This difference is far above the estimated uncertainties. 
In this case, the magnetic tongues mostly add a constant shift to the estimated tilt, producing a tilt angle that opposes the Joy's law prediction. 
This effect is mostly removed with CoFFE, achieving a better estimation of the FR axis intrinsic tilt at all analysed times.  
The computed difference between $\phia$ and $\phic$ is also above the mean dispersion reported in several Joy's law studies \citep[\eg\ ][]{Wang15,McClintock16}. 
Similarly to what  we have seen in the previous examples, there is a strong apparent rotation of $\approx11\degree$ in the counter-clockwise sense due to the evolution of the magnetic tongues, while CoFFE provides an almost constant tilt along the same period of time.

\subsection{Summary of the AR results} 
\label{sect_summary}

In Table~\ref{tableta}, we compare the different tilt-angle parameters obtained from both estimations, magnetic barycentres, and CoFFE {with $p=0$}, for the analysed ARs. The four ARs were selected as representative bipolar regions showing strong tongues along their full observed emergence. 
Columns one and two show the AR NOAA number and the twist sign identified from the magnetic tongues as done by \citet{Luoni11}.
In columns three and four we compare the mean tilt angles, apparent and derived from CoFFE, computed along the AR emergence 
and noted $\overline{\phi}_a$ and $\overline{\phic}$, respectively.
In three cases, ARs 9574, 9906, and 10268, magnetic tongues affect the field distribution producing values of {the apparent tilt} $\phia$ such that the sign of $\overline{\phia}$ is opposite to Joy's law (Figures \ref{fig_9574}c, \ref{fig_10268}c and \ref{fig_9906}c).
In contrast, the estimation of $\phic$ is in accordance with Joy's law along most of the AR emergence. 

Column five in Table~\ref{tableta} shows the mean difference between both estimations [$\overline{|\phia -\phic|}$].
There is a significant difference between $\phia$ and $\phic$, especially in the aforementioned cases the mean correction achieved with $\phic$ is above $9\degree$. 
It is noteworthy that the largest correction corresponds to AR 10268, a case in which the tongue flux is more entangled to the core, and therefore, represents the most difficult scenario expected for the optimal performance of CoFFE. 

In columns six and seven we show the mean tilt variations per day [$\Delta \phia$] and [$\Delta \phic$], respectively.
The tilt variation per day is defined positive (negative) when it corresponds to a counter-clockwise (clockwise) rotation of the bipole. 
{We chose to show these two quantities to highlight the role of magnetic tongues on the dispersion observed in studies that determine Joy's law using daily-averaged tilts taking few data per day and at random times \citep[{\eg}][]{Li12}.}

In previous studies, the rotation of the bipole is used as a proxy of the FR writhe. In all the analysed ARs, the magnetic tongues produced strong spurious rotations, detected in the estimation of $\phia$ during the first day of the emergence. 
Depending on the AR inclination and the sign of the twist the corrected rotation obtained with $\phic$ can be enhanced (ARs 9574 and 9748) or diminished (ARs 9906 and 10268) by the presence of the magnetic tongues. 

All the examples have a significant difference between $\phic$ and $\phia$.
 Even the case with one iteration ($\nit=1$) provides typically a good approximation of $\phic$ , except for the initial phase of the AR emergence.
For all the analysed cases, a fast convergence is achieved with a mean number of iterations between $4$ and $6$ (see {right} column in Table~\ref{tableta}), though there are a few isolated magnetograms for which $\Cit$ is larger than 10.
This parameter indicates how fast the method converges to $\phic$ in average and it has a direct impact on the performance of the code. 
With a mean number of iterations of $5$, the procedure takes less than ten seconds per magnetogram to compute $\phic$ in a regular desktop computer. 

For all the examples, we find a reliable correction of the tilt angle for $p=0$ and {the number of iterations to achieve convergence ($\nit=\Cit$)} because we are able to correct the spurious rotations produced by the enhanced flux of the magnetic tongues. 
The field distribution in the ARs shows a clearer separation between the tongue and the core field than in the analytical models, therefore, in most cases, we find that $p=0$ is enough to obtain a good approximation of $\phii$, thus removing substantially the effect of the tongues. 
Despite the increase of the uncertainties, larger $p$ values also provide a consistent estimation of $\phic$, but as $p$ increases the computations have a tendency to have larger fluctuations.

\section{Summary and conclusions}
 \label{sect_Conclusions} 

The estimation of the tilt angle of solar bipolar ARs using LOS magnetograms is affected by the intrinsic properties of the emerging FR. 
In this work, we have focused on the effect of the FR twist on the photospheric magnetic field distribution during the emergence of ARs.
The evolution of the magnetic tongues shifts the location of the flux-weighted centre of the polarities (or magnetic barycentres), which can be interpreted as an apparent tilt [$\phia$] or rotation of the bipole. 
We develop a novel method named Core Field Fit Estimator (CoFFE) that partially removes the aforementioned effects and provides a corrected estimation of the tilt angle. 

The procedure is based on the identification of two different magnetic flux distributions, defined as core and tongue, that best represent the contribution of the axial and the azimuthal field projection in the LOS direction.
We estimate the core field centres of each polarity using a Gaussian fit of the deprojected magnetograms and we compute the corrected tilt angle [$\phic$] {with CoFFE}. 
The procedure is simplified to have only one input parameter, called $p$, which defines the reduction of the fitted area in each magnetogram removing most of the tongue field contribution from the tilt estimation.

In \sect{CM}, we tested CoFFE using an analytical model based on the kinematic rise of a half-torus FR  \citep[see][]{Poisson16}. 
Using this model we are able to study the evolution of the axial and the azimuthal fields separately and to compare them with the core and tongue locations provided by CoFFE. 
The tilt correction achieved for $p=0$ is around $45\%$ of the difference between $\phia$ and $\phii$ along the FR emergence, being this correction {nearly} independent of the twist amplitude (see \figs{TILTIT}{tilt-n}). However, for the case of non-uniform twist with $h=1$, the effective correction provided by $\phic$ is reduced to a range between $25 \%$ to $50\%$ of $\phia$ (see \fig{tilt-h}). 
Although the use of $p = 1$ removes up to $85\%$ of the core magnetic flux, the value of $\phic$ for this parameter provides an improved estimation of the intrinsic tilt [$\phii$] along all the FR emergence. 
For the uniform twist models the correction is around $60\%$, while for the model with a non-uniform radial twist profile {increasing towards the FR border} ($h=1$) the correction is slightly lower; the difference between $\phic$ and $\phii$ is reduced to approximately $55\%$ of $\phia$.
Therefore, for all the models tested with different twist strengths, we achieved a better approximation to the intrinsic tilt of the FR axis than the one given by $\phia$.

In \sect{Correction}, we studied the emergence of four bipolar ARs using SOHO/MDI magnetograms to compute $\phic$ using CoFFE. 
We selected these ARs because their magnetic tongues are strong along the full emergence. 
In the studied cases the magnetic barycentres are strongly shifted towards the tongues producing a large variation on the measurements of {the apparent tilt} $\phia$. 
For all the analysed magnetograms, we found that CoFFE achieved a clear identification of both the core and tongue dominant fields.
The removal of the tongue field from the tilt estimation allows us to obtain an estimation of $\phic$ which is not only compatible with the Coriolis force action in each hemisphere (Joy's law) but also presumably closer to the FR intrinsic tilt.
 { It is noteworthy that the assumed Gaussian profile, used for the core region of observed ARs, is highly idealised and can be a limitation to the strength of the method. Although it is out of the scope of the present work, we intend to explore the influence of the core profile on the intrinsic properties derived from CoFFE in the future.}

The core and tongue fields are not easily separable in the synthetic magnetograms obtained from the FR model because of the continuous transition between the two fields.  
In contrast, ARs have magnetic tongues which are at least partially detached from the core field.
This kind of field distribution was predicted by the MHD simulations of \citet{Archontis10} in which the azimuthal field projected on the vertical direction presented concentrated structures, described as fingers. These structures appeared due to the presence of transverse converging plasma flows on the photosphere.
The aforementioned characteristic of the tongue field, observed in the analysed ARs, improved the performance of our procedure to identify the core centre position and distribution. 
Consequently, we found that CoFFE is {stable and weakly dependent of the selection of the parameter $p$ within the interval [-1,1]} when applied to ARs. 
We found that a value of $p=0$ is sufficient to obtain a good estimation of $\phic$ along the AR emergence. 
As less flux is removed from the fit, than for cases with $p>0$, the results {provide} a more {robust fit of the core centres}.
Therefore, CoFFE can be used automatically to correct the tongue effect independently of the AR studied. 

In three of the cases (ARs 9574, 10268, and 9906), the departure of the apparent tilt angle {$\phia$} from Joy's law was larger than the mean deviation {and even of opposite sign than the tilt} reported in several statistical studies \citep[\eg\ ][]{Wang15,Li12,McClintock14}. 
This implies that the magnetic tongue effect over the tilt angle can impact the variation and the mean deviation in the determination of Joy's law. 
Although an statistical analysis is still needed to support this statement, it is noteworthy that the effect of the magnetic tongues is not considered in observational studies that use daily measurements of the tilt angle of ARs. 

The correct determination of the tilt angle is essential for understanding the complex mechanisms involved in magnetic flux emergence. 
{ The method proposed in this work provides a more precise estimation of this parameter using AR observations of the magnetic field projected along the line-of-sight. {In a similar way, it can be applied to the vertical component of the field in} vector magnetograms.  
}
We consider that CoFFE is an important new complement to several observational studies that rely on the precise determination of the tilt angle, in particular by allowing for the analysis to be extended to the early stage of AR emergence. 
     
%
 \begin{acknowledgements}

The authors are grateful for the insightful referee who stimulated us to improve several parts of the manuscript.
MP, MLF, and CHM acknowledge financial support from Argentine grants PICT 2012-0973 (ANPCyT), UBACyT 20020130100321 and PIP 2012-01-403 (CONICET). MLF and CHM are members of the Carrera del Investigador Cient\'{\i}fico of the Consejo Nacional de Investigaciones Cient\'{\i}ficas y T\'ecnicas (CONICET). MP is a fellow of CONICET. {This work was supported by the Programme National PNST of CNRS/INSU co-funded by CNES and CEA.} The authors acknowledge the use of data from the SOHO (ESA/NASA) mission. These data are produced by the MDI international consortia.

 \end{acknowledgements}
 
\begin{appendix} 
\section{CoFFE applied to non-uniform twist models} \label{app_extra}

\begin{figure}[!h]
\begin{center}
\includegraphics[width=.4\textwidth]{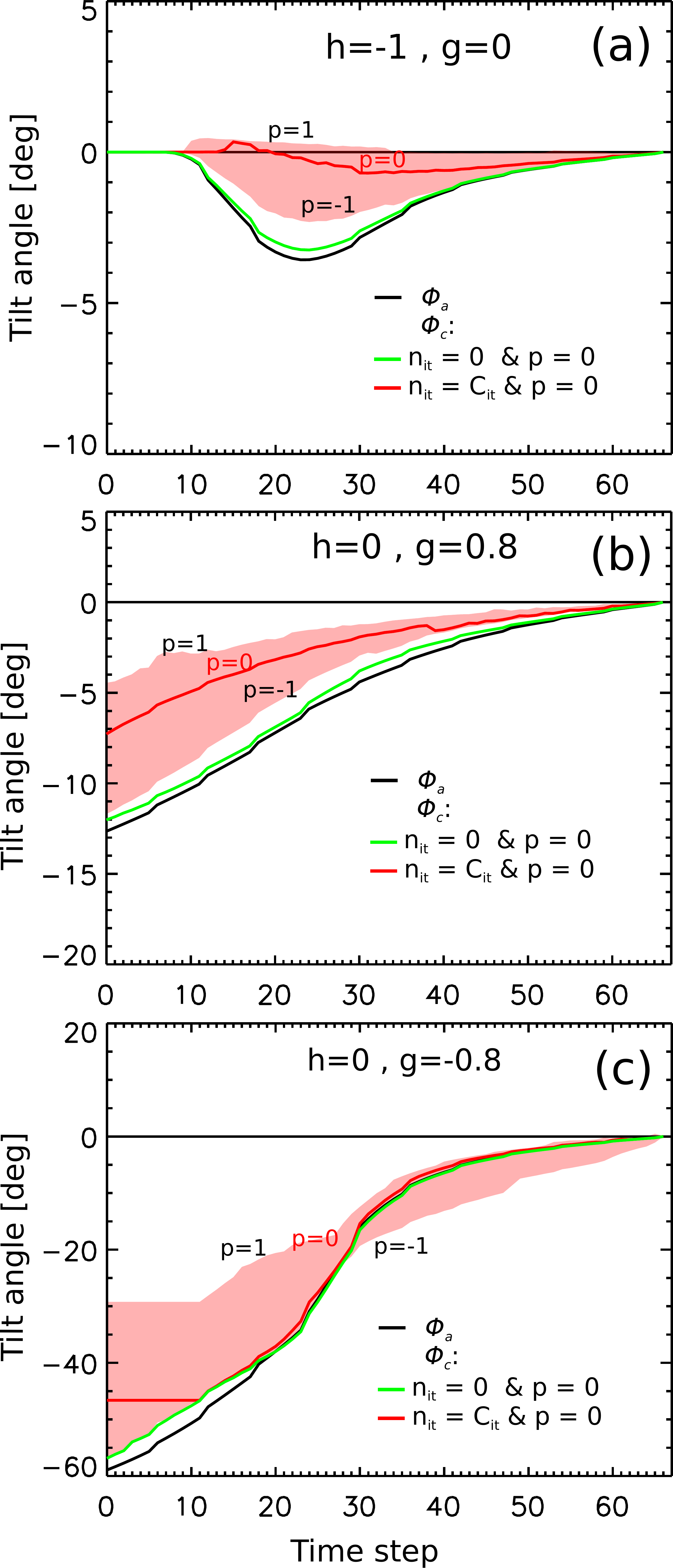}
\caption{Evolution of the tilt angle for non-uniform twist models with {${\Nto } = 0.5$ and (a) $h=-1$, $g = 0$,  (b) $h=0$, $g=0.8$, and (c) $h=0$, $g=-0.8$.}
The black line shows the {apparent} tilt angle estimations obtained from the polarity barycentres, $\phia$. 
We plot $\phic$ computed with CoFFE without (green) and with convergence (red) achieved (see inset).
The red-shaded area corresponds to the estimation of $\phic$ computed with different values of $p$ within the interval $[-1,1]$ {and after achieving the convergence criterion of the tilt angle described in \sect{GM}}.
The associated movies are available online (fig11\_a.avi, fig11\_b.avi, and fig11\_c.avi).
}
 \label{fig_tilt-appendix}
\end{center} 
\end{figure}

{ In this appendix, we further scan the space of parameters of the flux rope model with a non-uniform twist. We show the results for three pairs of $h$ and $g$ values, complementing the other extreme pair shown in \fig{tilt-h}. 

The model with $h=-1$ and $g=0$ has a strong reduction of the twist towards the edges of the FR, {\eq{Nt}}, so the effect of the tongues is weak along the full emergence (see movie ``fig11\_a.avi''). 
The correction achieved with CoFFE is close to the intrinsic tilt $\phii$ ($p=0$, red line in \fig{tilt-appendix}a). 

The model with $h=0$ and $g=0.8$, \fig{tilt-appendix}b, has a tilt angle which 
departs from $\phii$ in a similar way as the one obtained from a uniform twist model (\figs{TILTIT}{tilt-n}). 
Then, the results of CoFFE are also similar (\fig{tilt-appendix}b).
 
Finally, the case with $h=0$ and $g=-0.8$ is an extreme case (\fig{tilt-appendix}c). 
In this model the field lines concentrate more at the bottom of the FR than at the top, producing a strong asymmetry between the inner and the outer part of each polarity \citep{Poisson16}. 
The resulting $B_{z}$ is less concentrated at the core centre position but stronger towards the PIL. 
Therefore, the polarity barycentres are shifted towards the centre of the AR, closer to the magnetic tongues, and $\phia$ increases along the emergence  (see ``fig11\_c.avi'').  As seen in the case with $h=1$ (\fig{tilt-h}), a partial correction of the tilt angle can be achieved using $p>1$ (\fig{tilt-appendix}c). 
}

\end{appendix}

\bibliographystyle{aa}  
\bibliography{paper_tongues}  
\IfFileExists{\jobname.bbl}{}
{\typeout{}
\typeout{****************************************************}
\typeout{****************************************************}
\typeout{** Please run "bibtex \jobname" to obtain}
\typeout{** the bibliography and then re-run LaTeX}
\typeout{** twice to fix the references!}
\typeout{****************************************************}
\typeout{****************************************************}
\typeout{}
}

\end{document}